\pdfoutput=1

\documentclass[copyright,creativecommons]{eptcs}

\hypersetup{pdfborder=0 0 0.5}
\hypersetup{pdfborderstyle={}}

\makeatletter
\@ifundefined{lhs2tex.lhs2tex.sty.read}%
  {\@namedef{lhs2tex.lhs2tex.sty.read}{}%
   \newcommand\SkipToFmtEnd{}%
   \newcommand\EndFmtInput{}%
   \long\def\SkipToFmtEnd#1\EndFmtInput{}%
  }\SkipToFmtEnd

\newcommand\ReadOnlyOnce[1]{\@ifundefined{#1}{\@namedef{#1}{}}\SkipToFmtEnd}
\usepackage{amstext}
\usepackage{amssymb}
\usepackage{stmaryrd}
\DeclareFontFamily{OT1}{cmtex}{}
\DeclareFontShape{OT1}{cmtex}{m}{n}
  {<5><6><7><8>cmtex8
   <9>cmtex9
   <10><10.95><12><14.4><17.28><20.74><24.88>cmtex10}{}
\DeclareFontShape{OT1}{cmtex}{m}{it}
  {<-> ssub * cmtt/m/it}{}

\DeclareFontShape{OT1}{cmtt}{bx}{n}
  {<5><6><7><8>cmtt8
   <9>cmbtt9
   <10><10.95><12><14.4><17.28><20.74><24.88>cmbtt10}{}
\DeclareFontShape{OT1}{cmtex}{bx}{n}
  {<-> ssub * cmtt/bx/n}{}

\newcommand{\Conid}[1]{\mathit{#1}}
\newcommand{\Varid}[1]{\mathit{#1}}
\newcommand{\anonymous}{\kern0.06em \vbox{\hrule\@width.5em}}

\newcommand{\bind}{\mathbin{>\!\!\!>\mkern-6.7mu=}}

\usepackage{polytable}

\@ifundefined{mathindent}%
  {\newdimen\mathindent\mathindent\leftmargini}%
  {}%

\def\resethooks{%
  \global\let\SaveRestoreHook\empty
  \global\let\ColumnHook\empty}
\newcommand*{\savecolumns}[1][default]%
  {\g@addto@macro\SaveRestoreHook{\savecolumns[#1]}}
\newcommand*{\restorecolumns}[1][default]%
  {\g@addto@macro\SaveRestoreHook{\restorecolumns[#1]}}
\newcommand*{\aligncolumn}[2]%
  {\g@addto@macro\ColumnHook{\column{#1}{#2}}}

\resethooks

\newcommand{\onelinecommentchars}{\quad-{}- }
\newcommand{\commentbeginchars}{\enskip\{-}
\newcommand{\commentendchars}{-\}\enskip}

\newcommand{\visiblecomments}{%
  \let\onelinecomment=\onelinecommentchars
  \let\commentbegin=\commentbeginchars
  \let\commentend=\commentendchars}

\newcommand{\invisiblecomments}{%
  \let\onelinecomment=\empty
  \let\commentbegin=\empty
  \let\commentend=\empty}

\visiblecomments

\newlength{\blanklineskip}
\newcommand{\hsindent}[1]{\quad}%
\let\hspre\empty
\let\hspost\empty

\EndFmtInput
\makeatother
\ReadOnlyOnce{polycode.fmt}%
\makeatletter

\newcommand{\hsnewpar}[1]%
  {{\parskip=0pt\parindent=0pt\par\vskip #1\noindent}}

\newcommand{\hscodestyle}{}

\newcommand{\sethscode}[1]%
  {\expandafter\let\expandafter\hscode\csname #1\endcsname
   \expandafter\let\expandafter\endhscode\csname end#1\endcsname}

  {\par\noindent
   \advance\leftskip\mathindent
   \hscodestyle
   \let\\=\@normalcr
   \(\pboxed}%
  {\endpboxed\)%
   \par\noindent
   \ignorespacesafterend}

  {\hsnewpar\abovedisplayskip
   \advance\leftskip\mathindent
   \hscodestyle
   \let\hspre\(\let\hspost\)%
   \pboxed}%
  {\endpboxed%
   \hsnewpar\belowdisplayskip
   \ignorespacesafterend}

  {\hsnewpar\abovedisplayskip
   \advance\leftskip\mathindent
   \hscodestyle
   \let\\=\@normalcr
   \(\pboxed}%
  {\endpboxed\)%
   \hsnewpar\belowdisplayskip
   \ignorespacesafterend}

\newcommand{\plainhs}{\sethscode{plainhscode}}

\plainhs

  {\hsnewpar\abovedisplayskip
   \advance\leftskip\mathindent
   \hscodestyle
   \let\\=\@normalcr
   \(\parray}%
  {\endparray\)%
   \hsnewpar\belowdisplayskip
   \ignorespacesafterend}

  {\parray}{\endparray}

  {\(\parray}{\endparray\)}

\def\codeframewidth{\arrayrulewidth}
\RequirePackage{calc}

  {\parskip=\abovedisplayskip\par\noindent
   \hscodestyle
   \arrayrulewidth=\codeframewidth
   \tabular{@{}|p{\linewidth-2\arraycolsep-2\arrayrulewidth-2pt}|@{}}%
   \hline\framedhslinecorrect\\{-1.5ex}%
   \let\endoflinesave=\\
   \let\\=\@normalcr
   \(\pboxed}%
  {\endpboxed\)%
   \framedhslinecorrect\endoflinesave{.5ex}\hline
   \endtabular
   \parskip=\belowdisplayskip\par\noindent
   \ignorespacesafterend}

\newcommand{\framedhslinecorrect}[2]%
  {#1[#2]}

  {\(\def\column##1##2{}%
   \let\>\undefined\let\<\undefined\let\\\undefined
   \newcommand\>[1][]{}\newcommand\<[1][]{}\newcommand\\[1][]{}%
   \def\fromto##1##2##3{##3}%
   }{\) }%

  {\let\orighscode=\hscode
   \let\origendhscode=\endhscode
   \def\endhscode{\def\hscode{\endgroup\def\@currenvir{hscode}\\}\begingroup}
   \orighscode\def\hscode{\endgroup\def\@currenvir{hscode}}}%
  {\origendhscode
   \global\let\hscode=\orighscode
   \global\let\endhscode=\origendhscode}%

\makeatother
\EndFmtInput
\newcommand{\constructor}[1]{\mathsf{#1}}

\usepackage{braket}
\usepackage{amsmath}
\usepackage{amssymb}
\usepackage{amsfonts}
\usepackage{stmaryrd}
\usepackage{bbm}

\newcommand{\doi}[1]{doi:\href{http://dx.doi.org/#1}{%
    \urlstyle{same}\nolinkurl{#1}}}
\usepackage{natbib}
\setcitestyle{authoryear,round,semicolon,aysep={},yysep={,}}

\usepackage[english]{babel}
\usepackage[mathletters]{ucs}
\usepackage[utf8x]{inputenc}
\usepackage{autofe}

\DeclareUnicodeCharacter{"2115}{\ensuremath{\mathbb{N}}}
\DeclareUnicodeCharacter{"211D}{\ensuremath{\mathbb{R}}}
\DeclareUnicodeCharacter{"2237}{\ensuremath{::}}
\DeclareUnicodeCharacter{"27E6}{\ensuremath{\llbracket}}
\DeclareUnicodeCharacter{"27E7}{\ensuremath{\rrbracket}}

\begin{document}

\title{Beating the Productivity Checker\\Using Embedded Languages}
\newcommand{\titlerunning}{Beating the Productivity Checker Using Embedded Languages}
\author{Nils Anders Danielsson\institute{University of Nottingham}}
\newcommand{\authorrunning}{Nils Anders Danielsson}
\hypersetup{pdftitle={\titlerunning},pdfauthor={\authorrunning}}

\maketitle

\begin{abstract}
  Some total languages, like Agda and Coq, allow the use of guarded
  corecursion to construct infinite values and proofs. Guarded
  corecursion is a form of recursion in which arbitrary recursive
  calls are allowed, as long as they are guarded by a coinductive
  constructor. Guardedness ensures that programs are productive, i.e.
  that every finite prefix of an infinite value can be computed in
  finite time. However, many productive programs are not guarded, and
  it can be nontrivial to put them in guarded form.

  This paper gives a method for turning a productive program into a
  guarded program. The method amounts to defining a problem-specific
  language as a data type, writing the program in the problem-specific
  language, and writing a guarded interpreter for this language.
\end{abstract}

\section{Introduction}
\label{sec:introduction}

When working with infinite values in a total setting it is common to
require that every value is \emph{productive} \citep{sijtsma}: even
though a value is conceptually infinite, it should always be possible
to compute the next unit of information in finite time.
The primitive methods for defining infinite values in the proof
assistants Agda and Coq are based on \emph{guarded corecursion}
\citep{coquand}, which is a conservative approximation of
productivity for coinductive types.
The basic idea of guarded corecursion is that ``corecursive calls''
may only take place under guarding constructors, thus ensuring that
the next unit of information---the next constructor---can always be
computed.
For instance, consider the following definition of \ensuremath{\Varid{nats_{≥}}\;\Varid{n}}, the
stream of successive natural numbers greater than or equal to \ensuremath{\Varid{n}}
(\ensuremath{{\anonymous \mkern1mu\constructor{∷}\mkern-1mu\anonymous }} is the cons constructor for streams):
\begin{hscode}\SaveRestoreHook
\column{B}{@{}>{\hspre}l<{\hspost}@{}}%
\column{3}{@{}>{\hspre}l<{\hspost}@{}}%
\column{E}{@{}>{\hspre}l<{\hspost}@{}}%
\>[3]{}\Varid{nats_{≥}}\;\mathbin{:}\;\Conid{ℕ}\;\Varid{→}\;\Conid{Stream}\;\Conid{ℕ}{}\<[E]%
\\
\>[3]{}\Varid{nats_{≥}}\;\Varid{n}\;\mathrel{=}\;\Varid{n}\;\constructor{∷}\;\Varid{nats_{≥}}\;(\constructor{suc}\;\Varid{n}){}\<[E]%
\ColumnHook
\end{hscode}\resethooks
This definition is guarded, and has the property that the next natural
number can always be computed in finite time.
As another example, consider \ensuremath{\Varid{bad}}:
\begin{hscode}\SaveRestoreHook
\column{B}{@{}>{\hspre}l<{\hspost}@{}}%
\column{3}{@{}>{\hspre}l<{\hspost}@{}}%
\column{E}{@{}>{\hspre}l<{\hspost}@{}}%
\>[3]{}\Varid{bad}\;\mathbin{:}\;\Conid{Stream}\;\Conid{ℕ}{}\<[E]%
\\
\>[3]{}\Varid{bad}\;\mathrel{=}\;\Varid{tail}\;(\constructor{zero}\;\constructor{∷}\;\Varid{bad}){}\<[E]%
\ColumnHook
\end{hscode}\resethooks
This ``definition'' is not guarded (due to the presence of \ensuremath{\Varid{tail}}),
nor is it productive: \ensuremath{\Varid{bad}} is not well-defined.
Finally consider the following definition of the stream of natural
numbers:
\begin{hscode}\SaveRestoreHook
\column{B}{@{}>{\hspre}l<{\hspost}@{}}%
\column{3}{@{}>{\hspre}l<{\hspost}@{}}%
\column{E}{@{}>{\hspre}l<{\hspost}@{}}%
\>[3]{}\Varid{nats}\;\mathbin{:}\;\Conid{Stream}\;\Conid{ℕ}{}\<[E]%
\\
\>[3]{}\Varid{nats}\;\mathrel{=}\;\constructor{zero}\;\constructor{∷}\;\Varid{map}\;\constructor{suc}\;\Varid{nats}{}\<[E]%
\ColumnHook
\end{hscode}\resethooks
This definition is productive, but unfortunately it is not guarded,
because \ensuremath{\Varid{map}} is not a constructor. In fact, many productive
definitions are not guarded, and it can be nontrivial to find
equivalent guarded definitions.

The main contribution of this paper is a technique for translating a
large class of productive but unguarded definitions into guarded
definitions.
The basic observation of the technique is that many productive
definitions would be guarded if some functions were actually
constructors. For instance, if \ensuremath{\Varid{map}} were a constructor, then \ensuremath{\Varid{nats}}
would be guarded.
The technique then amounts to defining a problem-specific language as
a data type which includes a constructor for every function like
\ensuremath{\Varid{map}}, implementing the productive definitions in a guarded way using
this language, and implementing a guarded interpreter for the
language.
Optionally one can also prove that the resulting definitions satisfy
their intended defining equations, and that these equations have
unique solutions.

The technique relies on the use of data types defined using mixed
induction and coinduction (see Section~\ref{sec:coinduction}), so it
requires a programming language with support for such definitions.
The examples in the paper have been implemented using Agda
\citep{norell,agda}, a dependently typed, total\footnote{Agda is an
  experimental system with neither a formalised meta-theory nor a
  verified type checker, so take words such as ``total'' with a grain
  of salt.} functional programming language with good support for
mixed induction and coinduction.
The supporting source code is available to download
\citep{accompanying-code-par-2010}.

Before we continue it may be useful to state some things which are
\emph{not} addressed by the paper:
\begin{itemize}
\item The paper's focus is on establishing productivity, not on
  representing non-productive definitions, nor on making
  non-productive definitions total by restricting their types
  \citep{bertot}.
\item No attempt is made to automate the technique: as it stands it
  provides a manual, somewhat ad hoc method for getting productive
  definitions accepted by a system based on guarded corecursion.
\end{itemize}

The rest of the paper is structured as follows:
Section~\ref{sec:coinduction} discusses induction and coinduction in
the context of Agda,
Sections~\ref{sec:programs}--\ref{sec:nested-applications} (as well as
Appendix~\ref{sec:inductive-stream-equality}) introduce the
language-based approach to productivity through a number of examples,
Section~\ref{sec:related-work} discusses related work, and
Section~\ref{sec:conclusions} concludes.

\section{Mixed Induction and Coinduction}
\label{sec:coinduction}

This section gives a quick introduction to Agda, in particular to its
support for mixed induction and coinduction. For more details, see
\citet[Section~2]{danielsson-altenkirch-subtyping}.

In Agda the type of infinite streams can be defined as follows:
\begin{hscode}\SaveRestoreHook
\column{B}{@{}>{\hspre}l<{\hspost}@{}}%
\column{3}{@{}>{\hspre}l<{\hspost}@{}}%
\column{5}{@{}>{\hspre}l<{\hspost}@{}}%
\column{E}{@{}>{\hspre}l<{\hspost}@{}}%
\>[3]{}\mathbf{data}\;\Conid{Stream}\;(\Conid{A}\;\mathbin{:}\;\Conid{Set})\;\mathbin{:}\;\Conid{Set}\;\mathbf{where}{}\<[E]%
\\
\>[3]{}\hsindent{2}{}\<[5]%
\>[5]{}{\anonymous \mkern1mu\constructor{∷}\mkern-1mu\anonymous }\;\mathbin{:}\;\Conid{A}\;\Varid{→}\;\Varid{∞}\;(\Conid{Stream}\;\Conid{A})\;\Varid{→}\;\Conid{Stream}\;\Conid{A}{}\<[E]%
\ColumnHook
\end{hscode}\resethooks
This definition states that \ensuremath{\Conid{Stream}\;\Conid{A}} is a \ensuremath{\Conid{Set}} (``type'') with a
single (infix) constructor \ensuremath{{\anonymous \mkern1mu\constructor{∷}\mkern-1mu\anonymous }} of type\linebreak\ensuremath{\Conid{A}\;\Varid{→}\;\Varid{∞}\;(\Conid{Stream}\;\Conid{A})\;\Varid{→}\;\Conid{Stream}\;\Conid{A}}.
The inclusion of \ensuremath{\Varid{∞}} in the type of \ensuremath{{\anonymous \mkern1mu\constructor{∷}\mkern-1mu\anonymous }} makes \ensuremath{\Conid{Stream}\;\Conid{A}}
coinductive; without it the type would be empty.
You should read \ensuremath{\Varid{∞}\;(\Conid{Stream}\;\Conid{A})} as ``delayed stream of \ensuremath{\Conid{A}}s''---the
function \ensuremath{\Varid{∞}\;\mathbin{:}\;\Conid{Set}\;\Varid{→}\;\Conid{Set}} is analogous to the suspension type
constructors which are sometimes used to introduce non-strictness in
strict languages \citep{wadler-taha-macqueen}, and closely related to
the domain-theoretic notion of lifting.
However, Agda programs are required to be total.

We can construct infinite values by guarded corecursion. For instance,
we can define a function which combines two streams in a pointwise
manner as follows:\footnote{The notation \ensuremath{\{\mkern-1mu\Conid{A}\;\Conid{B}\;\Conid{C}\;\mathbin{:}\;\Conid{Set}\mkern-1mu\}\;\Varid{→}\;\Varid{…}} means
  that \ensuremath{\Varid{zip\mkern-2mu{}With}} takes three \emph{implicit} arguments \ensuremath{\Conid{A}}, \ensuremath{\Conid{B}} and
  \ensuremath{\Conid{C}}, all of type \ensuremath{\Conid{Set}}. These arguments do not need to be given
  explicitly if Agda can infer them.}
\begin{hscode}\SaveRestoreHook
\column{B}{@{}>{\hspre}l<{\hspost}@{}}%
\column{3}{@{}>{\hspre}l<{\hspost}@{}}%
\column{E}{@{}>{\hspre}l<{\hspost}@{}}%
\>[3]{}\Varid{zip\mkern-2mu{}With}\;\mathbin{:}\;\{\mkern-1mu\Conid{A}\;\Conid{B}\;\Conid{C}\;\mathbin{:}\;\Conid{Set}\mkern-1mu\}\;\Varid{→}\;(\Conid{A}\;\Varid{→}\;\Conid{B}\;\Varid{→}\;\Conid{C})\;\Varid{→}\;\Conid{Stream}\;\Conid{A}\;\Varid{→}\;\Conid{Stream}\;\Conid{B}\;\Varid{→}\;\Conid{Stream}\;\Conid{C}{}\<[E]%
\\
\>[3]{}\Varid{zip\mkern-2mu{}With}\;\Varid{f}\;(\Varid{x}\;\constructor{∷}\;\Varid{xs})\;(\Varid{y}\;\constructor{∷}\;\Varid{ys})\;\mathrel{=}\;\Varid{f}\;\Varid{x}\;\Varid{y}\;\constructor{∷}\;\raisebox{0.8ex}[2ex]{\footnotesize ♯}\;\Varid{zip\mkern-2mu{}With}\;\Varid{f}\;(\raisebox{0.8ex}[2ex]{\footnotesize ♭}\;\Varid{xs})\;(\raisebox{0.8ex}[2ex]{\footnotesize ♭}\;\Varid{ys}){}\<[E]%
\ColumnHook
\end{hscode}\resethooks
This definition uses the coinductive delay constructor \ensuremath{\raisebox{0.8ex}[2ex]{\footnotesize ♯}\mkern-2mu\anonymous }
(sharp)\footnote{The prefix operator \ensuremath{\raisebox{0.8ex}[2ex]{\footnotesize ♯}\mkern-2mu\anonymous } is the most tightly binding
  operator in this paper; ordinary function application binds tighter,
  though.} and the force function \ensuremath{\raisebox{0.8ex}[2ex]{\footnotesize ♭}} (flat):
\begin{hscode}\SaveRestoreHook
\column{B}{@{}>{\hspre}l<{\hspost}@{}}%
\column{5}{@{}>{\hspre}l<{\hspost}@{}}%
\column{9}{@{}>{\hspre}l<{\hspost}@{}}%
\column{26}{@{}>{\hspre}l<{\hspost}@{}}%
\column{33}{@{}>{\hspre}l<{\hspost}@{}}%
\column{E}{@{}>{\hspre}l<{\hspost}@{}}%
\>[5]{}\raisebox{0.8ex}[2ex]{\footnotesize ♯}\mkern-2mu\anonymous \;{}\<[9]%
\>[9]{}\mathbin{:}\;\{\mkern-1mu\Conid{A}\;\mathbin{:}\;\Conid{Set}\mkern-1mu\}\;\Varid{→}\;{}\<[26]%
\>[26]{}\Conid{A}\;\Varid{→}\;\Varid{∞}\;{}\<[33]%
\>[33]{}\Conid{A}{}\<[E]%
\\
\>[5]{}\raisebox{0.8ex}[2ex]{\footnotesize ♭}\;{}\<[9]%
\>[9]{}\mathbin{:}\;\{\mkern-1mu\Conid{A}\;\mathbin{:}\;\Conid{Set}\mkern-1mu\}\;\Varid{→}\;\Varid{∞}\;{}\<[26]%
\>[26]{}\Conid{A}\;\Varid{→}\;{}\<[33]%
\>[33]{}\Conid{A}{}\<[E]%
\ColumnHook
\end{hscode}\resethooks
Agda views \ensuremath{\Varid{zip\mkern-2mu{}With}} as guarded, because there is no non-constructor
function between the left-hand side and the corecursive call, and
there is at least one use of the guarding coinductive constructor
\ensuremath{\raisebox{0.8ex}[2ex]{\footnotesize ♯}\mkern-2mu\anonymous }.
This constructor has special status: it is treated as a constructor by
Agda's productivity checker, but may not be used in patterns. Instead
one can use the force function: \ensuremath{\raisebox{0.8ex}[2ex]{\footnotesize ♭}\;(\raisebox{0.8ex}[2ex]{\footnotesize ♯}\;\Varid{x})} reduces to \ensuremath{\Varid{x}}.

As another example, consider the following definition of
equality---bisimilarity---for streams (which makes use of the fact
that constructors can be overloaded):
\begin{hscode}\SaveRestoreHook
\column{B}{@{}>{\hspre}l<{\hspost}@{}}%
\column{3}{@{}>{\hspre}l<{\hspost}@{}}%
\column{5}{@{}>{\hspre}l<{\hspost}@{}}%
\column{E}{@{}>{\hspre}l<{\hspost}@{}}%
\>[3]{}\mathbf{data}\;{\anonymous \mkern-4mu≈\mkern-6mu\anonymous }\;\{\mkern-1mu\Conid{A}\;\mathbin{:}\;\Conid{Set}\mkern-1mu\}\;\mathbin{:}\;\Conid{Stream}\;\Conid{A}\;\Varid{→}\;\Conid{Stream}\;\Conid{A}\;\Varid{→}\;\Conid{Set}\;\mathbf{where}{}\<[E]%
\\
\>[3]{}\hsindent{2}{}\<[5]%
\>[5]{}{\anonymous \mkern1mu\constructor{∷}\mkern-1mu\anonymous }\;\mathbin{:}\;(\Varid{x}\;\mathbin{:}\;\Conid{A})\;\Varid{→}\;\{\mkern-1mu\Varid{xs}\;\Varid{ys}\;\mathbin{:}\;\Varid{∞}\;(\Conid{Stream}\;\Conid{A})\mkern-1mu\}\;\Varid{→}\;\Varid{∞}\;(\raisebox{0.8ex}[2ex]{\footnotesize ♭}\;\Varid{xs}\;≈\;\raisebox{0.8ex}[2ex]{\footnotesize ♭}\;\Varid{ys})\;\Varid{→}\;\Varid{x}\;\constructor{∷}\;\Varid{xs}\;≈\;\Varid{x}\;\constructor{∷}\;\Varid{ys}{}\<[E]%
\ColumnHook
\end{hscode}\resethooks
This definition states that two streams are equal if their heads are
identical and their tails are equal (coinductively).
Note that the elements of this type are equality \emph{proofs}.
We can establish equalities by constructing proofs using guarded
corecursion. For instance, we can prove symmetry as follows:
\begin{hscode}\SaveRestoreHook
\column{B}{@{}>{\hspre}l<{\hspost}@{}}%
\column{3}{@{}>{\hspre}l<{\hspost}@{}}%
\column{E}{@{}>{\hspre}l<{\hspost}@{}}%
\>[3]{}\Varid{sym}\;\mathbin{:}\;\{\mkern-1mu\Conid{A}\;\mathbin{:}\;\Conid{Set}\mkern-1mu\}\;\Varid{→}\;\{\mkern-1mu\Varid{xs}\;\Varid{ys}\;\mathbin{:}\;\Conid{Stream}\;\Conid{A}\mkern-1mu\}\;\Varid{→}\;\Varid{xs}\;≈\;\Varid{ys}\;\Varid{→}\;\Varid{ys}\;≈\;\Varid{xs}{}\<[E]%
\\
\>[3]{}\Varid{sym}\;(\Varid{x}\;\constructor{∷}\;\Varid{xs}\mkern-5mu≈\mkern-6mu\Varid{ys})\;\mathrel{=}\;\Varid{x}\;\constructor{∷}\;\raisebox{0.8ex}[2ex]{\footnotesize ♯}\;\Varid{sym}\;(\raisebox{0.8ex}[2ex]{\footnotesize ♭}\;\Varid{xs}\mkern-5mu≈\mkern-6mu\Varid{ys}){}\<[E]%
\ColumnHook
\end{hscode}\resethooks
(Note that \ensuremath{\Varid{xs}\mkern-5mu≈\mkern-6mu\Varid{ys}} is an ordinary variable, albeit perhaps with an
unusual name.)

Let us now consider a definition which uses both induction and
coinduction.
The type \ensuremath{\Conid{SP}\;\Conid{A}\;\Conid{B}} of stream processors
\citep{hancock-et-al}---representations of programs taking streams of
\ensuremath{\Conid{A}}s to streams of \ensuremath{\Conid{B}}s---can be defined as follows:
\begin{hscode}\SaveRestoreHook
\column{B}{@{}>{\hspre}l<{\hspost}@{}}%
\column{3}{@{}>{\hspre}l<{\hspost}@{}}%
\column{5}{@{}>{\hspre}l<{\hspost}@{}}%
\column{10}{@{}>{\hspre}l<{\hspost}@{}}%
\column{28}{@{}>{\hspre}l<{\hspost}@{}}%
\column{E}{@{}>{\hspre}l<{\hspost}@{}}%
\>[3]{}\mathbf{data}\;\Conid{SP}\;(\Conid{A}\;\Conid{B}\;\mathbin{:}\;\Conid{Set})\;\mathbin{:}\;\Conid{Set}\;\mathbf{where}{}\<[E]%
\\
\>[3]{}\hsindent{2}{}\<[5]%
\>[5]{}\constructor{put}\;{}\<[10]%
\>[10]{}\mathbin{:}\;\Conid{B}\;\Varid{→}\;\Varid{∞}\;(\Conid{SP}\;\Conid{A}\;\Conid{B})\;{}\<[28]%
\>[28]{}\Varid{→}\;\Conid{SP}\;\Conid{A}\;\Conid{B}{}\<[E]%
\\
\>[3]{}\hsindent{2}{}\<[5]%
\>[5]{}\constructor{get}\;{}\<[10]%
\>[10]{}\mathbin{:}\;(\Conid{A}\;\Varid{→}\;\Conid{SP}\;\Conid{A}\;\Conid{B})\;{}\<[28]%
\>[28]{}\Varid{→}\;\Conid{SP}\;\Conid{A}\;\Conid{B}{}\<[E]%
\ColumnHook
\end{hscode}\resethooks
Here \ensuremath{\constructor{put}\;\Varid{b}\;\Varid{sp}} is intended to output \ensuremath{\Varid{b}} and continue with \ensuremath{\Varid{sp}},
while \ensuremath{\constructor{get}\;\Varid{f}} is intended to read an element \ensuremath{\Varid{a}} and continue with \ensuremath{\Varid{f}\;\Varid{a}}.
You can see the type as the nested fixpoint\footnote{Currently this is
  not quite correct in Agda \citep{altenkirch-danielsson}, but for the
  purposes of this paper the differences are irrelevant.} \ensuremath{\Varid{νX.}\;\Varid{μY.}\;\Conid{B}\;\Varid{×}\;\Conid{X}\;\Varid{+}\;(\Conid{A}\;\Varid{→}\;\Conid{Y})}---in fact, \emph{all} (non-mutual) data types in the
paper can be seen as nested fixpoints of the form \ensuremath{\Varid{νX.}\;\Varid{μY.}\;\Conid{F}\;\Conid{X}\;\Conid{Y}} (and
mutually defined data types can be merged by adding an index).
Note that the recursive argument of \ensuremath{\constructor{put}} is delayed (coinductive),
whereas the recursive argument of \ensuremath{\constructor{get}} is not.
This means that we can have an infinite number of consecutive \ensuremath{\constructor{put}}
constructors, but only a finite number of consecutive \ensuremath{\constructor{get}}s;
definitions such as the following one are not guarded and not
accepted:
\begin{hscode}\SaveRestoreHook
\column{B}{@{}>{\hspre}l<{\hspost}@{}}%
\column{3}{@{}>{\hspre}l<{\hspost}@{}}%
\column{E}{@{}>{\hspre}l<{\hspost}@{}}%
\>[3]{}\Varid{sink}\;\mathbin{:}\;\{\mkern-1mu\Conid{A}\;\Conid{B}\;\mathbin{:}\;\Conid{Set}\mkern-1mu\}\;\Varid{→}\;\Conid{SP}\;\Conid{A}\;\Conid{B}{}\<[E]%
\\
\>[3]{}\Varid{sink}\;\mathrel{=}\;\constructor{get}\;(\lambda\;\anonymous \;\Varid{→}\;\Varid{sink}){}\<[E]%
\ColumnHook
\end{hscode}\resethooks
The definition of \ensuremath{\Varid{sink}} is not problematic in and of itself (assuming
that it is not evaluated too eagerly). However, by ruling out such
definitions we make other definitions possible, for instance the
following one, which gives the semantics of a stream processor:
\begin{hscode}\SaveRestoreHook
\column{B}{@{}>{\hspre}l<{\hspost}@{}}%
\column{3}{@{}>{\hspre}l<{\hspost}@{}}%
\column{15}{@{}>{\hspre}l<{\hspost}@{}}%
\column{27}{@{}>{\hspre}l<{\hspost}@{}}%
\column{E}{@{}>{\hspre}l<{\hspost}@{}}%
\>[3]{}\Varid{⟦}\mkern-1mu\anonymous \mkern1mu\Varid{⟧}\;\mathbin{:}\;\{\mkern-1mu\Conid{A}\;\Conid{B}\;\mathbin{:}\;\Conid{Set}\mkern-1mu\}\;\Varid{→}\;\Conid{SP}\;\Conid{A}\;\Conid{B}\;\Varid{→}\;\Conid{Stream}\;\Conid{A}\;\Varid{→}\;\Conid{Stream}\;\Conid{B}{}\<[E]%
\\
\>[3]{}\Varid{⟦}\;\constructor{put}\;\Varid{b}\;\Varid{sp}\;{}\<[15]%
\>[15]{}\Varid{⟧}\;\Varid{as}\;{}\<[27]%
\>[27]{}\mathrel{=}\;\Varid{b}\;\constructor{∷}\;\raisebox{0.8ex}[2ex]{\footnotesize ♯}\;(\Varid{⟦}\;\raisebox{0.8ex}[2ex]{\footnotesize ♭}\;\Varid{sp}\;\Varid{⟧}\;\Varid{as}){}\<[E]%
\\
\>[3]{}\Varid{⟦}\;\constructor{get}\;\Varid{f}\;{}\<[15]%
\>[15]{}\Varid{⟧}\;(\Varid{a}\;\constructor{∷}\;\Varid{as})\;{}\<[27]%
\>[27]{}\mathrel{=}\;\Varid{⟦}\;\Varid{f}\;\Varid{a}\;\Varid{⟧}\;(\raisebox{0.8ex}[2ex]{\footnotesize ♭}\;\Varid{as}){}\<[E]%
\ColumnHook
\end{hscode}\resethooks
This function is accepted by Agda because it is defined using a
lexicographic combination of guarded corecursion and structural
recursion. In this particular example the first component of the
lexicographic product is the ``guardedness'', and the second component
is the \emph{inductive} structure of the stream processor:
\begin{itemize}
\item In the first clause the corecursive call is guarded. The stream
  processor is not structurally smaller, due to the use of the force
  function (\ensuremath{\raisebox{0.8ex}[2ex]{\footnotesize ♭}}), but this is irrelevant.
\item In the second clause the corecursive call is not guarded, but
  there is no non-constructor function between the left-hand side and
  the corecursive call, so we say that ``guardedness is preserved''.
  On the other hand, the stream processor argument is strictly
  structurally smaller (\ensuremath{\Varid{f}\;\Varid{x}} is smaller than \ensuremath{\constructor{get}\;\Varid{f}} for any \ensuremath{\Varid{x}}).
\end{itemize}
Armed with the knowledge that there can only be a finite number of
consecutive \ensuremath{\constructor{get}} constructors we conclude that, when evaluating \ensuremath{\Varid{⟦}\;\Varid{sp}\;\Varid{⟧}\;\Varid{as}}, we must eventually reach the first clause. At this stage we can
immediately inspect the head element of the output stream, because the
second clause does not introduce any interfering destructors.

As a final example, consider \ensuremath{\Varid{filter}}, which is not accepted by Agda:
\begin{hscode}\SaveRestoreHook
\column{B}{@{}>{\hspre}l<{\hspost}@{}}%
\column{3}{@{}>{\hspre}l<{\hspost}@{}}%
\column{30}{@{}>{\hspre}l<{\hspost}@{}}%
\column{E}{@{}>{\hspre}l<{\hspost}@{}}%
\>[3]{}\Varid{filter}\;\mathbin{:}\;\{\mkern-1mu\Conid{A}\;\mathbin{:}\;\Conid{Set}\mkern-1mu\}\;\Varid{→}\;(\Conid{A}\;\Varid{→}\;\Conid{Bool})\;\Varid{→}\;\Conid{Stream}\;\Conid{A}\;\Varid{→}\;\Conid{Stream}\;\Conid{A}{}\<[E]%
\\
\>[3]{}\Varid{filter}\;\Varid{p}\;(\Varid{x}\;\constructor{∷}\;\Varid{xs})\;\mathbf{with}\;\Varid{p}\;\Varid{x}{}\<[E]%
\\
\>[3]{}\Varid{filter}\;\Varid{p}\;(\Varid{x}\;\constructor{∷}\;\Varid{xs})\;\mid \;\constructor{true}\;{}\<[30]%
\>[30]{}\mathrel{=}\;\Varid{x}\;\constructor{∷}\;\raisebox{0.8ex}[2ex]{\footnotesize ♯}\;\Varid{filter}\;\Varid{p}\;(\raisebox{0.8ex}[2ex]{\footnotesize ♭}\;\Varid{xs}){}\<[E]%
\\
\>[3]{}\Varid{filter}\;\Varid{p}\;(\Varid{x}\;\constructor{∷}\;\Varid{xs})\;\mid \;\constructor{false}\;{}\<[30]%
\>[30]{}\mathrel{=}\;\Varid{filter}\;\Varid{p}\;(\raisebox{0.8ex}[2ex]{\footnotesize ♭}\;\Varid{xs}){}\<[E]%
\ColumnHook
\end{hscode}\resethooks
(Here the \ensuremath{\mathbf{with}} construct is used to pattern match on \mbox{\ensuremath{\Varid{p}\;\Varid{x}}}.)
The first corecursive call is guarded, but in the last clause the call
is not guarded, and nothing is structurally smaller, so this function
is not accepted.

The explanations above should suffice to understand the definitions in
this paper---in fact, most definitions use less complicated recursion
principles than the one used by \ensuremath{\Varid{⟦}\mkern-1mu\anonymous \mkern1mu\Varid{⟧}}.
For more information about Agda's criterion for accepting a function
as total, see \citet[Section~2.5]{danielsson-altenkirch-subtyping}.

Before we continue note that, in order to reduce clutter, the
declarations of implicit arguments have been omitted in the remainder
of the paper.

\section{Making Programs Guarded}
\label{sec:programs}

As noted in the introduction guardedness is sometimes an inconvenient
restriction: there are productive programs which are not syntactically
guarded. This section introduces a language-based technique for
putting definitions in guarded form.

Consider the following definition of the stream of Fibonacci numbers:
\begin{hscode}\SaveRestoreHook
\column{B}{@{}>{\hspre}l<{\hspost}@{}}%
\column{3}{@{}>{\hspre}l<{\hspost}@{}}%
\column{E}{@{}>{\hspre}l<{\hspost}@{}}%
\>[3]{}\Varid{fib}\;\mathbin{:}\;\Conid{Stream}\;\Conid{ℕ}{}\<[E]%
\\
\>[3]{}\Varid{fib}\;\mathrel{=}\;0\;\constructor{∷}\;\raisebox{0.8ex}[2ex]{\footnotesize ♯}\;\Varid{zip\mkern-2mu{}With}\;{\anonymous \mkern-0.5mu\Varid{+}\mkern-2mu\anonymous \mkern3mu}\;\Varid{fib}\;(1\;\constructor{∷}\;\raisebox{0.8ex}[2ex]{\footnotesize ♯}\;\Varid{fib}){}\<[E]%
\ColumnHook
\end{hscode}\resethooks
While the definition of \ensuremath{\Varid{fib}} is productive, it is not guarded,
because the function \ensuremath{\Varid{zip\mkern-2mu{}With}} is not a constructor. If \ensuremath{\Varid{zip\mkern-2mu{}With}} were
a constructor the definition would be guarded, though, and this
presents a way out: we can define a problem-specific language which
includes \ensuremath{\Varid{zip\mkern-2mu{}With}} as a constructor, and then define an interpreter
for the language by using guarded corecursion.

A simple language of stream programs can be defined as
follows:\footnote{\ensuremath{\Conid{Set}_{\text{1}}} is a type of large types; \ensuremath{\Varid{∞}} has type \ensuremath{\Conid{Set}_{\Varid{i}}\;\Varid{→}\;\Conid{Set}_{\Varid{i}}} for any \ensuremath{\Varid{i}}.}
\begin{hscode}\SaveRestoreHook
\column{B}{@{}>{\hspre}l<{\hspost}@{}}%
\column{3}{@{}>{\hspre}l<{\hspost}@{}}%
\column{5}{@{}>{\hspre}l<{\hspost}@{}}%
\column{15}{@{}>{\hspre}l<{\hspost}@{}}%
\column{18}{@{}>{\hspre}l<{\hspost}@{}}%
\column{E}{@{}>{\hspre}l<{\hspost}@{}}%
\>[3]{}\mathbf{data}\;\Conid{Stream}_{\text{P}}\;\mathbin{:}\;\Conid{Set}\;\Varid{→}\;\Conid{Set}_{\text{1}}\;\mathbf{where}{}\<[E]%
\\
\>[3]{}\hsindent{2}{}\<[5]%
\>[5]{}{\anonymous \mkern1mu\constructor{∷}\mkern-1mu\anonymous }\;{}\<[15]%
\>[15]{}\mathbin{:}\;{}\<[18]%
\>[18]{}\Conid{A}\;\Varid{→}\;\Varid{∞}\;(\Conid{Stream}_{\text{P}}\;\Conid{A})\;\Varid{→}\;\Conid{Stream}_{\text{P}}\;\Conid{A}{}\<[E]%
\\
\>[3]{}\hsindent{2}{}\<[5]%
\>[5]{}\constructor{zip\mkern-0.8mu With}\;{}\<[15]%
\>[15]{}\mathbin{:}\;{}\<[18]%
\>[18]{}(\Conid{A}\;\Varid{→}\;\Conid{B}\;\Varid{→}\;\Conid{C})\;\Varid{→}\;\Conid{Stream}_{\text{P}}\;\Conid{A}\;\Varid{→}\;\Conid{Stream}_{\text{P}}\;\Conid{B}\;\Varid{→}\;\Conid{Stream}_{\text{P}}\;\Conid{C}{}\<[E]%
\ColumnHook
\end{hscode}\resethooks
Note that the stream program argument of \ensuremath{{\anonymous \mkern1mu\constructor{∷}\mkern-1mu\anonymous }} is coinductive, while
the arguments of \ensuremath{\constructor{zip\mkern-0.8mu With}} are inductive; a stream program consisting
of an infinitely deep application of \ensuremath{\constructor{zip\mkern-0.8mu With}}s would not be
productive.

Stream programs will be turned into streams in two steps. First a kind
of weak head normal form (WHNF) for stream programs is computed
recursively, and then the resulting stream is computed corecursively.
The WHNFs are defined in the following way:
\begin{hscode}\SaveRestoreHook
\column{B}{@{}>{\hspre}l<{\hspost}@{}}%
\column{3}{@{}>{\hspre}l<{\hspost}@{}}%
\column{5}{@{}>{\hspre}l<{\hspost}@{}}%
\column{E}{@{}>{\hspre}l<{\hspost}@{}}%
\>[3]{}\mathbf{data}\;\Conid{Stream}_{\text{W}}\;\mathbin{:}\;\Conid{Set}\;\Varid{→}\;\Conid{Set}_{\text{1}}\;\mathbf{where}{}\<[E]%
\\
\>[3]{}\hsindent{2}{}\<[5]%
\>[5]{}{\anonymous \mkern1mu\constructor{∷}\mkern-1mu\anonymous }\;\mathbin{:}\;\Conid{A}\;\Varid{→}\;\Conid{Stream}_{\text{P}}\;\Conid{A}\;\Varid{→}\;\Conid{Stream}_{\text{W}}\;\Conid{A}{}\<[E]%
\ColumnHook
\end{hscode}\resethooks
Note that the stream argument to \ensuremath{{\anonymous \mkern1mu\constructor{∷}\mkern-1mu\anonymous }} is a (``suspended'') program,
not a WHNF.
The function \ensuremath{\Varid{whnf}} which computes WHNFs can be defined by structural
recursion:
\begin{hscode}\SaveRestoreHook
\column{B}{@{}>{\hspre}l<{\hspost}@{}}%
\column{5}{@{}>{\hspre}l<{\hspost}@{}}%
\column{30}{@{}>{\hspre}l<{\hspost}@{}}%
\column{E}{@{}>{\hspre}l<{\hspost}@{}}%
\>[5]{}\Varid{whnf}\;\mathbin{:}\;\Conid{Stream}_{\text{P}}\;\Conid{A}\;\Varid{→}\;\Conid{Stream}_{\text{W}}\;\Conid{A}{}\<[E]%
\\
\>[5]{}\Varid{whnf}\;(\Varid{x}\;\constructor{∷}\;\Varid{xs})\;{}\<[30]%
\>[30]{}\mathrel{=}\;\Varid{x}\;\constructor{∷}\;\raisebox{0.8ex}[2ex]{\footnotesize ♭}\;\Varid{xs}{}\<[E]%
\\
\>[5]{}\Varid{whnf}\;(\constructor{zip\mkern-0.8mu With}\;\Varid{f}\;\Varid{xs}\;\Varid{ys})\;{}\<[30]%
\>[30]{}\mathrel{=}\;\Varid{zip\mkern-2mu{}With}_{\text{W}}\;\Varid{f}\;(\Varid{whnf}\;\Varid{xs})\;(\Varid{whnf}\;\Varid{ys}){}\<[E]%
\ColumnHook
\end{hscode}\resethooks
Here \ensuremath{\Varid{zip\mkern-2mu{}With}_{\text{W}}} is defined by simple case analysis:
\begin{hscode}\SaveRestoreHook
\column{B}{@{}>{\hspre}l<{\hspost}@{}}%
\column{5}{@{}>{\hspre}l<{\hspost}@{}}%
\column{E}{@{}>{\hspre}l<{\hspost}@{}}%
\>[5]{}\Varid{zip\mkern-2mu{}With}_{\text{W}}\;\mathbin{:}\;(\Conid{A}\;\Varid{→}\;\Conid{B}\;\Varid{→}\;\Conid{C})\;\Varid{→}\;\Conid{Stream}_{\text{W}}\;\Conid{A}\;\Varid{→}\;\Conid{Stream}_{\text{W}}\;\Conid{B}\;\Varid{→}\;\Conid{Stream}_{\text{W}}\;\Conid{C}{}\<[E]%
\\
\>[5]{}\Varid{zip\mkern-2mu{}With}_{\text{W}}\;\Varid{f}\;(\Varid{x}\;\constructor{∷}\;\Varid{xs})\;(\Varid{y}\;\constructor{∷}\;\Varid{ys})\;\mathrel{=}\;\Varid{f}\;\Varid{x}\;\Varid{y}\;\constructor{∷}\;\constructor{zip\mkern-0.8mu With}\;\Varid{f}\;\Varid{xs}\;\Varid{ys}{}\<[E]%
\ColumnHook
\end{hscode}\resethooks
WHNFs can then be turned into streams corecursively:
\begin{hscode}\SaveRestoreHook
\column{B}{@{}>{\hspre}l<{\hspost}@{}}%
\column{3}{@{}>{\hspre}l<{\hspost}@{}}%
\column{5}{@{}>{\hspre}l<{\hspost}@{}}%
\column{E}{@{}>{\hspre}l<{\hspost}@{}}%
\>[3]{}\mathbf{mutual}{}\<[E]%
\\
\>[3]{}\hsindent{2}{}\<[5]%
\>[5]{}\Varid{⟦}\mkern-1mu\anonymous \mkern1mu\Varid{⟧}_{\text{W}}\;\mathbin{:}\;\Conid{Stream}_{\text{W}}\;\Conid{A}\;\Varid{→}\;\Conid{Stream}\;\Conid{A}{}\<[E]%
\\
\>[3]{}\hsindent{2}{}\<[5]%
\>[5]{}\Varid{⟦}\;\Varid{x}\;\constructor{∷}\;\Varid{xs}\;\Varid{⟧}_{\text{W}}\;\mathrel{=}\;\Varid{x}\;\constructor{∷}\;\raisebox{0.8ex}[2ex]{\footnotesize ♯}\;\Varid{⟦}\;\Varid{xs}\;\Varid{⟧}_{\text{P}}{}\<[E]%
\\[\blanklineskip]%
\>[3]{}\hsindent{2}{}\<[5]%
\>[5]{}\Varid{⟦}\mkern-1mu\anonymous \mkern1mu\Varid{⟧}_{\text{P}}\;\mathbin{:}\;\Conid{Stream}_{\text{P}}\;\Conid{A}\;\Varid{→}\;\Conid{Stream}\;\Conid{A}{}\<[E]%
\\
\>[3]{}\hsindent{2}{}\<[5]%
\>[5]{}\Varid{⟦}\;\Varid{xs}\;\Varid{⟧}_{\text{P}}\;\mathrel{=}\;\Varid{⟦}\;\Varid{whnf}\;\Varid{xs}\;\Varid{⟧}_{\text{W}}{}\<[E]%
\ColumnHook
\end{hscode}\resethooks
Note that this definition is guarded. (Agda accepts definitions like
this one even though it is split up over two mutually defined
functions; alternatively one could write \ensuremath{\Varid{⟦}\;\Varid{x}\;\constructor{∷}\;\Varid{xs}\;\Varid{⟧}_{\text{W}}\;\mathrel{=}\;\Varid{x}\;\constructor{∷}\;\raisebox{0.8ex}[2ex]{\footnotesize ♯}\;\Varid{⟦}\;\Varid{whnf}\;\Varid{xs}\;\Varid{⟧}_{\text{W}}} and define \ensuremath{\Varid{⟦}\mkern-1mu\anonymous \mkern1mu\Varid{⟧}_{\text{P}}} separately.)

Given the language above we can now define the stream of Fibonacci
numbers using guarded corecursion:\\
\begin{minipage}{0.5\linewidth}
  \begin{hscode}\SaveRestoreHook
\column{B}{@{}>{\hspre}l<{\hspost}@{}}%
\column{5}{@{}>{\hspre}l<{\hspost}@{}}%
\column{E}{@{}>{\hspre}l<{\hspost}@{}}%
\>[5]{}\Varid{fib}_{\text{P}}\;\mathbin{:}\;\Conid{Stream}_{\text{P}}\;\Conid{ℕ}{}\<[E]%
\\
\>[5]{}\Varid{fib}_{\text{P}}\;\mathrel{=}\;0\;\constructor{∷}\;\raisebox{0.8ex}[2ex]{\footnotesize ♯}\;\constructor{zip\mkern-0.8mu With}\;{\anonymous \mkern-0.5mu\Varid{+}\mkern-2mu\anonymous \mkern3mu}\;\Varid{fib}_{\text{P}}\;(1\;\constructor{∷}\;\raisebox{0.8ex}[2ex]{\footnotesize ♯}\;\Varid{fib}_{\text{P}}){}\<[E]%
\ColumnHook
\end{hscode}\resethooks
\end{minipage}
\begin{minipage}{0.49\linewidth}
  \begin{hscode}\SaveRestoreHook
\column{B}{@{}>{\hspre}l<{\hspost}@{}}%
\column{5}{@{}>{\hspre}l<{\hspost}@{}}%
\column{E}{@{}>{\hspre}l<{\hspost}@{}}%
\>[5]{}\Varid{fib}\;\mathbin{:}\;\Conid{Stream}\;\Conid{ℕ}{}\<[E]%
\\
\>[5]{}\Varid{fib}\;\mathrel{=}\;\Varid{⟦}\;\Varid{fib}_{\text{P}}\;\Varid{⟧}_{\text{P}}{}\<[E]%
\ColumnHook
\end{hscode}\resethooks
\end{minipage}
One can prove that this definition satisfies the original equation for
\ensuremath{\Varid{fib}} by first proving corecursively that \ensuremath{\Varid{⟦}\mkern-1mu\anonymous \mkern1mu\Varid{⟧}_{\text{P}}} is homomorphic with
respect to \ensuremath{\Varid{zip\mkern-2mu{}With}}/\ensuremath{\constructor{zip\mkern-0.8mu With}}:
\begin{hscode}\SaveRestoreHook
\column{B}{@{}>{\hspre}l<{\hspost}@{}}%
\column{3}{@{}>{\hspre}l<{\hspost}@{}}%
\column{16}{@{}>{\hspre}l<{\hspost}@{}}%
\column{19}{@{}>{\hspre}l<{\hspost}@{}}%
\column{E}{@{}>{\hspre}l<{\hspost}@{}}%
\>[3]{}\Varid{zip\mkern-2mu{}With}\mkern-1mu\textit{-hom}\;{}\<[16]%
\>[16]{}\mathbin{:}\;{}\<[19]%
\>[19]{}(\Varid{f}\;\mathbin{:}\;\Conid{A}\;\Varid{→}\;\Conid{B}\;\Varid{→}\;\Conid{C})\;\Varid{→}\;(\Varid{xs}\;\mathbin{:}\;\Conid{Stream}\;\Conid{A})\;\Varid{→}\;(\Varid{ys}\;\mathbin{:}\;\Conid{Stream}\;\Conid{B})\;\Varid{→}\;{}\<[E]%
\\
\>[19]{}\Varid{⟦}\;\constructor{zip\mkern-0.8mu With}\;\Varid{f}\;\Varid{xs}\;\Varid{ys}\;\Varid{⟧}_{\text{P}}\;≈\;\Varid{zip\mkern-2mu{}With}\;\Varid{f}\;\Varid{⟦}\;\Varid{xs}\;\Varid{⟧}_{\text{P}}\;\Varid{⟦}\;\Varid{ys}\;\Varid{⟧}_{\text{P}}{}\<[E]%
\\
\>[3]{}\textit{fib-correct}\;{}\<[16]%
\>[16]{}\mathbin{:}\;{}\<[19]%
\>[19]{}\Varid{fib}\;≈\;0\;\constructor{∷}\;\raisebox{0.8ex}[2ex]{\footnotesize ♯}\;\Varid{zip\mkern-2mu{}With}\;{\anonymous \mkern-0.5mu\Varid{+}\mkern-2mu\anonymous \mkern3mu}\;\Varid{fib}\;(1\;\constructor{∷}\;\raisebox{0.8ex}[2ex]{\footnotesize ♯}\;\Varid{fib}){}\<[E]%
\ColumnHook
\end{hscode}\resethooks
For the omitted proofs, see \citet{accompanying-code-par-2010}.
One may also want to establish that the original equation for \ensuremath{\Varid{fib}}
defines the stream completely, i.e.\@ that it has a \emph{unique}
solution.
For an explanation of how this can be done, see
Section~\ref{sec:proofs}.

It can be instructive to see what would happen if we tried to use the
method above to implement the ill-defined stream \ensuremath{\Varid{bad}} from the
introduction.
Defining the language and giving a ``definition'' for \ensuremath{\Varid{bad}} is
straightforward:\\
\begin{minipage}{0.5\linewidth}
\begin{hscode}\SaveRestoreHook
\column{B}{@{}>{\hspre}l<{\hspost}@{}}%
\column{5}{@{}>{\hspre}l<{\hspost}@{}}%
\column{7}{@{}>{\hspre}l<{\hspost}@{}}%
\column{14}{@{}>{\hspre}l<{\hspost}@{}}%
\column{E}{@{}>{\hspre}l<{\hspost}@{}}%
\>[5]{}\mathbf{data}\;\Conid{Stream}_{\text{P}}\;(\Conid{A}\;\mathbin{:}\;\Conid{Set})\;\mathbin{:}\;\Conid{Set}\;\mathbf{where}{}\<[E]%
\\
\>[5]{}\hsindent{2}{}\<[7]%
\>[7]{}{\anonymous \mkern1mu\constructor{∷}\mkern-1mu\anonymous }\;{}\<[14]%
\>[14]{}\mathbin{:}\;\Conid{A}\;\Varid{→}\;\Varid{∞}\;(\Conid{Stream}_{\text{P}}\;\Conid{A})\;\Varid{→}\;\Conid{Stream}_{\text{P}}\;\Conid{A}{}\<[E]%
\\
\>[5]{}\hsindent{2}{}\<[7]%
\>[7]{}\constructor{tail}\;{}\<[14]%
\>[14]{}\mathbin{:}\;\Conid{Stream}_{\text{P}}\;\Conid{A}\;\Varid{→}\;\Conid{Stream}_{\text{P}}\;\Conid{A}{}\<[E]%
\ColumnHook
\end{hscode}\resethooks
\end{minipage}
\begin{minipage}{0.49\linewidth}
\begin{hscode}\SaveRestoreHook
\column{B}{@{}>{\hspre}l<{\hspost}@{}}%
\column{5}{@{}>{\hspre}l<{\hspost}@{}}%
\column{E}{@{}>{\hspre}l<{\hspost}@{}}%
\>[5]{}\Varid{bad}\;\mathbin{:}\;\Conid{Stream}_{\text{P}}\;\Conid{ℕ}{}\<[E]%
\\
\>[5]{}\Varid{bad}\;\mathrel{=}\;\constructor{tail}\;(\constructor{zero}\;\constructor{∷}\;\raisebox{0.8ex}[2ex]{\footnotesize ♯}\;\Varid{bad}){}\<[E]%
\ColumnHook
\end{hscode}\resethooks
\end{minipage}
However, turning stream programs into streams becomes tricky. How
would \ensuremath{\Varid{tail}_{\text{W}}} be defined?\\
\begin{minipage}{0.5\linewidth}
\begin{hscode}\SaveRestoreHook
\column{B}{@{}>{\hspre}l<{\hspost}@{}}%
\column{5}{@{}>{\hspre}l<{\hspost}@{}}%
\column{7}{@{}>{\hspre}l<{\hspost}@{}}%
\column{E}{@{}>{\hspre}l<{\hspost}@{}}%
\>[5]{}\mathbf{data}\;\Conid{Stream}_{\text{W}}\;(\Conid{A}\;\mathbin{:}\;\Conid{Set})\;\mathbin{:}\;\Conid{Set}\;\mathbf{where}{}\<[E]%
\\
\>[5]{}\hsindent{2}{}\<[7]%
\>[7]{}{\anonymous \mkern1mu\constructor{∷}\mkern-1mu\anonymous }\;\mathbin{:}\;\Conid{A}\;\Varid{→}\;\Conid{Stream}_{\text{P}}\;\Conid{A}\;\Varid{→}\;\Conid{Stream}_{\text{W}}\;\Conid{A}{}\<[E]%
\ColumnHook
\end{hscode}\resethooks
\end{minipage}
\begin{minipage}{0.49\linewidth}
\begin{hscode}\SaveRestoreHook
\column{B}{@{}>{\hspre}l<{\hspost}@{}}%
\column{5}{@{}>{\hspre}l<{\hspost}@{}}%
\column{E}{@{}>{\hspre}l<{\hspost}@{}}%
\>[5]{}\Varid{tail}_{\text{W}}\;\mathbin{:}\;\Conid{Stream}_{\text{W}}\;\Conid{A}\;\Varid{→}\;\Conid{Stream}_{\text{W}}\;\Conid{A}{}\<[E]%
\\
\>[5]{}\Varid{tail}_{\text{W}}\;(\Varid{x}\;\constructor{∷}\;\Varid{xs})\;\mathrel{=}\;\Varid{?}{}\<[E]%
\ColumnHook
\end{hscode}\resethooks
\end{minipage}
Note that, in the body of \ensuremath{\Varid{tail}_{\text{W}}}, \ensuremath{\Varid{xs}} is a stream program, but we
need to produce a WHNF.

\section{Several Types at Once}
\label{sec:multiple-types}

The technique introduced in Section~\ref{sec:programs} is not limited
to streams. In fact, it can be used with several types at the same
time.
To illustrate how this can be done I will implement circular
breadth-first labelling of trees \'a la \citet{jones-gibbons}.

The following type of potentially infinite binary trees will be used:
\begin{hscode}\SaveRestoreHook
\column{B}{@{}>{\hspre}l<{\hspost}@{}}%
\column{3}{@{}>{\hspre}l<{\hspost}@{}}%
\column{5}{@{}>{\hspre}l<{\hspost}@{}}%
\column{11}{@{}>{\hspre}l<{\hspost}@{}}%
\column{E}{@{}>{\hspre}l<{\hspost}@{}}%
\>[3]{}\mathbf{data}\;\Conid{Tree}\;(\Conid{A}\;\mathbin{:}\;\Conid{Set})\;\mathbin{:}\;\Conid{Set}\;\mathbf{where}{}\<[E]%
\\
\>[3]{}\hsindent{2}{}\<[5]%
\>[5]{}\constructor{leaf}\;{}\<[11]%
\>[11]{}\mathbin{:}\;\Conid{Tree}\;\Conid{A}{}\<[E]%
\\
\>[3]{}\hsindent{2}{}\<[5]%
\>[5]{}\constructor{node}\;{}\<[11]%
\>[11]{}\mathbin{:}\;\Varid{∞}\;(\Conid{Tree}\;\Conid{A})\;\Varid{→}\;\Conid{A}\;\Varid{→}\;\Varid{∞}\;(\Conid{Tree}\;\Conid{A})\;\Varid{→}\;\Conid{Tree}\;\Conid{A}{}\<[E]%
\ColumnHook
\end{hscode}\resethooks
Jones and Gibbons' implementation can be described as follows. First a
labelling function \ensuremath{\Varid{lab}} is defined. This function takes a tree, along
with a stream of streams of new labels. The labels in a prefix of the
$n$-th stream are used to label the $n$-th level of the tree, and the
remaining labels are returned from \ensuremath{\Varid{lab}}:
\begin{hscode}\SaveRestoreHook
\column{B}{@{}>{\hspre}l<{\hspost}@{}}%
\column{5}{@{}>{\hspre}l<{\hspost}@{}}%
\column{7}{@{}>{\hspre}l<{\hspost}@{}}%
\column{12}{@{}>{\hspre}l<{\hspost}@{}}%
\column{21}{@{}>{\hspre}l<{\hspost}@{}}%
\column{23}{@{}>{\hspre}l<{\hspost}@{}}%
\column{41}{@{}>{\hspre}l<{\hspost}@{}}%
\column{68}{@{}>{\hspre}l<{\hspost}@{}}%
\column{E}{@{}>{\hspre}l<{\hspost}@{}}%
\>[5]{}\Varid{lab}\;\mathbin{:}\;\Conid{Tree}\;\Conid{A}\;\Varid{→}\;\Conid{Stream}\;(\Conid{Stream}\;\Conid{B})\;\Varid{→}\;\Conid{Tree}\;\Conid{B}\;\Varid{×}\;\Conid{Stream}\;(\Conid{Stream}\;\Conid{B}){}\<[E]%
\\
\>[5]{}\Varid{lab}\;\constructor{leaf}\;{}\<[23]%
\>[23]{}\Varid{bss}\;{}\<[41]%
\>[41]{}\mathrel{=}\;(\constructor{leaf},{}\<[68]%
\>[68]{}\Varid{bss}){}\<[E]%
\\
\>[5]{}\Varid{lab}\;(\constructor{node}\;\Varid{l}\;\anonymous \;\Varid{r})\;{}\<[23]%
\>[23]{}((\Varid{b}\;\constructor{∷}\;\Varid{bs})\;\constructor{∷}\;\Varid{bss})\;{}\<[41]%
\>[41]{}\mathrel{=}\;(\constructor{node}\;(\raisebox{0.8ex}[2ex]{\footnotesize ♯}\;\Varid{l′})\;\Varid{b}\;(\raisebox{0.8ex}[2ex]{\footnotesize ♯}\;\Varid{r′}),{}\<[68]%
\>[68]{}\raisebox{0.8ex}[2ex]{\footnotesize ♭}\;\Varid{bs}\;\constructor{∷}\;\raisebox{0.8ex}[2ex]{\footnotesize ♯}\;\Varid{bss″}){}\<[E]%
\\
\>[5]{}\hsindent{2}{}\<[7]%
\>[7]{}\mathbf{where}{}\<[E]%
\\
\>[5]{}\hsindent{2}{}\<[7]%
\>[7]{}(\Varid{l′}{}\<[12]%
\>[12]{},\Varid{bss′})\;{}\<[21]%
\>[21]{}\mathrel{=}\;\Varid{lab}\;(\raisebox{0.8ex}[2ex]{\footnotesize ♭}\;\Varid{l})\;(\raisebox{0.8ex}[2ex]{\footnotesize ♭}\;\Varid{bss}){}\<[E]%
\\
\>[5]{}\hsindent{2}{}\<[7]%
\>[7]{}(\Varid{r′}{}\<[12]%
\>[12]{},\Varid{bss″})\;{}\<[21]%
\>[21]{}\mathrel{=}\;\Varid{lab}\;(\raisebox{0.8ex}[2ex]{\footnotesize ♭}\;\Varid{r})\;\Varid{bss′}{}\<[E]%
\ColumnHook
\end{hscode}\resethooks
This code is not accepted by Agda, because the recursive calls are not
guarded (their results are destructed, and furthermore \ensuremath{\Varid{lab}}, which is
not a constructor, is applied to a part of one of the
results).\footnote{Agda does not support pattern matching in where
  clauses as used here, but one can use projection functions instead.}
The next step in Jones and Gibbons' implementation is to construct the
stream of streams of labels which is used by \ensuremath{\Varid{lab}}, and use these
streams to compute the relabelled tree. This is done using a circular
construction:
\begin{hscode}\SaveRestoreHook
\column{B}{@{}>{\hspre}l<{\hspost}@{}}%
\column{5}{@{}>{\hspre}l<{\hspost}@{}}%
\column{7}{@{}>{\hspre}l<{\hspost}@{}}%
\column{E}{@{}>{\hspre}l<{\hspost}@{}}%
\>[5]{}\Varid{label}\;\mathbin{:}\;\Conid{Tree}\;\Conid{A}\;\Varid{→}\;\Conid{Stream}\;\Conid{B}\;\Varid{→}\;\Conid{Tree}\;\Conid{B}{}\<[E]%
\\
\>[5]{}\Varid{label}\;\Varid{t}\;\Varid{bs}\;\mathrel{=}\;\Varid{t′}{}\<[E]%
\\
\>[5]{}\hsindent{2}{}\<[7]%
\>[7]{}\mathbf{where}\;(\Varid{t′},\Varid{bss})\;\mathrel{=}\;\Varid{lab}\;\Varid{t}\;(\Varid{bs}\;\constructor{∷}\;\raisebox{0.8ex}[2ex]{\footnotesize ♯}\;\Varid{bss}){}\<[E]%
\ColumnHook
\end{hscode}\resethooks
This code is not accepted by Agda, because \ensuremath{\Varid{lab}} is not a constructor,
and furthermore the result of \ensuremath{\Varid{lab}} is destructed.

To implement breadth-first labelling in the style of Jones and Gibbons
the following universe of trees, streams, products and arbitrary
(small) types will be used:\\
\begin{minipage}{0.5\linewidth}
  \begin{hscode}\SaveRestoreHook
\column{B}{@{}>{\hspre}l<{\hspost}@{}}%
\column{3}{@{}>{\hspre}l<{\hspost}@{}}%
\column{5}{@{}>{\hspre}l<{\hspost}@{}}%
\column{13}{@{}>{\hspre}l<{\hspost}@{}}%
\column{E}{@{}>{\hspre}l<{\hspost}@{}}%
\>[3]{}\mathbf{data}\;\Conid{U}\;\mathbin{:}\;\Conid{Set}_{\text{1}}\;\mathbf{where}{}\<[E]%
\\
\>[3]{}\hsindent{2}{}\<[5]%
\>[5]{}\constructor{tree}\;{}\<[13]%
\>[13]{}\mathbin{:}\;\Conid{U}\;\Varid{→}\;\Conid{U}{}\<[E]%
\\
\>[3]{}\hsindent{2}{}\<[5]%
\>[5]{}\constructor{stream}\;{}\<[13]%
\>[13]{}\mathbin{:}\;\Conid{U}\;\Varid{→}\;\Conid{U}{}\<[E]%
\\
\>[3]{}\hsindent{2}{}\<[5]%
\>[5]{}{\anonymous \mkern1mu\Varid{⊗}\mkern-1mu\anonymous }\;{}\<[13]%
\>[13]{}\mathbin{:}\;\Conid{U}\;\Varid{→}\;\Conid{U}\;\Varid{→}\;\Conid{U}{}\<[E]%
\\
\>[3]{}\hsindent{2}{}\<[5]%
\>[5]{}{\Varid{⌈}\mkern-2mu\anonymous \mkern-0.5mu\Varid{⌉}}\;{}\<[13]%
\>[13]{}\mathbin{:}\;\Conid{Set}\;\Varid{→}\;\Conid{U}{}\<[E]%
\ColumnHook
\end{hscode}\resethooks
\end{minipage}
\begin{minipage}{0.49\linewidth}
  \begin{hscode}\SaveRestoreHook
\column{B}{@{}>{\hspre}l<{\hspost}@{}}%
\column{3}{@{}>{\hspre}l<{\hspost}@{}}%
\column{18}{@{}>{\hspre}l<{\hspost}@{}}%
\column{E}{@{}>{\hspre}l<{\hspost}@{}}%
\>[3]{}\Conid{El}\;\mathbin{:}\;\Conid{U}\;\Varid{→}\;\Conid{Set}{}\<[E]%
\\
\>[3]{}\Conid{El}\;(\constructor{tree}\;\Varid{a})\;{}\<[18]%
\>[18]{}\mathrel{=}\;\Conid{Tree}\;(\Conid{El}\;\Varid{a}){}\<[E]%
\\
\>[3]{}\Conid{El}\;(\constructor{stream}\;\Varid{a})\;{}\<[18]%
\>[18]{}\mathrel{=}\;\Conid{Stream}\;(\Conid{El}\;\Varid{a}){}\<[E]%
\\
\>[3]{}\Conid{El}\;(\Varid{a}\;\Varid{⊗}\;\Varid{b})\;{}\<[18]%
\>[18]{}\mathrel{=}\;\Conid{El}\;\Varid{a}\;\Varid{×}\;\Conid{El}\;\Varid{b}{}\<[E]%
\\
\>[3]{}\Conid{El}\;\Varid{⌈}\;\Conid{A}\;\Varid{⌉}\;{}\<[18]%
\>[18]{}\mathrel{=}\;\Conid{A}{}\<[E]%
\ColumnHook
\end{hscode}\resethooks
\end{minipage}
The type \ensuremath{\Conid{U}} defines codes for elements of the universe, and \ensuremath{\Conid{El}}
interprets these codes.

By indexing the program and WHNF types by codes from the universe \ensuremath{\Conid{U}}
we can work with several types at once:
\begin{hscode}\SaveRestoreHook
\column{B}{@{}>{\hspre}l<{\hspost}@{}}%
\column{3}{@{}>{\hspre}l<{\hspost}@{}}%
\column{5}{@{}>{\hspre}l<{\hspost}@{}}%
\column{7}{@{}>{\hspre}l<{\hspost}@{}}%
\column{13}{@{}>{\hspre}l<{\hspost}@{}}%
\column{E}{@{}>{\hspre}l<{\hspost}@{}}%
\>[3]{}\mathbf{mutual}{}\<[E]%
\\
\>[3]{}\hsindent{2}{}\<[5]%
\>[5]{}\mathbf{data}\;\Conid{El}_{\text{P}}\;\mathbin{:}\;\Conid{U}\;\Varid{→}\;\Conid{Set}_{\text{1}}\;\mathbf{where}{}\<[E]%
\\
\>[5]{}\hsindent{2}{}\<[7]%
\>[7]{}\Varid{↓}\;{}\<[13]%
\>[13]{}\mathbin{:}\;\Conid{El}_{\text{W}}\;\Varid{a}\;\Varid{→}\;\Conid{El}_{\text{P}}\;\Varid{a}{}\<[E]%
\\
\>[5]{}\hsindent{2}{}\<[7]%
\>[7]{}\constructor{fst}\;{}\<[13]%
\>[13]{}\mathbin{:}\;\Conid{El}_{\text{P}}\;(\Varid{a}\;\Varid{⊗}\;\Varid{b})\;\Varid{→}\;\Conid{El}_{\text{P}}\;\Varid{a}{}\<[E]%
\\
\>[5]{}\hsindent{2}{}\<[7]%
\>[7]{}\constructor{snd}\;{}\<[13]%
\>[13]{}\mathbin{:}\;\Conid{El}_{\text{P}}\;(\Varid{a}\;\Varid{⊗}\;\Varid{b})\;\Varid{→}\;\Conid{El}_{\text{P}}\;\Varid{b}{}\<[E]%
\\
\>[5]{}\hsindent{2}{}\<[7]%
\>[7]{}\constructor{lab}\;{}\<[13]%
\>[13]{}\mathbin{:}\;\Conid{Tree}\;\Conid{A}\;\Varid{→}\;\Conid{El}_{\text{P}}\;(\constructor{stream}\;\Varid{⌈}\;\Conid{Stream}\;\Conid{B}\;\Varid{⌉})\;\Varid{→}\;\Conid{El}_{\text{P}}\;(\constructor{tree}\;\Varid{⌈}\;\Conid{B}\;\Varid{⌉}\;\Varid{⊗}\;\constructor{stream}\;\Varid{⌈}\;\Conid{Stream}\;\Conid{B}\;\Varid{⌉}){}\<[E]%
\\[\blanklineskip]%
\>[3]{}\hsindent{2}{}\<[5]%
\>[5]{}\mathbf{data}\;\Conid{El}_{\text{W}}\;\mathbin{:}\;\Conid{U}\;\Varid{→}\;\Conid{Set}_{\text{1}}\;\mathbf{where}{}\<[E]%
\\
\>[5]{}\hsindent{2}{}\<[7]%
\>[7]{}\constructor{leaf}\;{}\<[13]%
\>[13]{}\mathbin{:}\;\Conid{El}_{\text{W}}\;(\constructor{tree}\;\Varid{a}){}\<[E]%
\\
\>[5]{}\hsindent{2}{}\<[7]%
\>[7]{}\constructor{node}\;{}\<[13]%
\>[13]{}\mathbin{:}\;\Varid{∞}\;(\Conid{El}_{\text{P}}\;(\constructor{tree}\;\Varid{a}))\;\Varid{→}\;\Conid{El}_{\text{W}}\;\Varid{a}\;\Varid{→}\;\Varid{∞}\;(\Conid{El}_{\text{P}}\;(\constructor{tree}\;\Varid{a}))\;\Varid{→}\;\Conid{El}_{\text{W}}\;(\constructor{tree}\;\Varid{a}){}\<[E]%
\\
\>[5]{}\hsindent{2}{}\<[7]%
\>[7]{}{\anonymous \mkern1mu\constructor{∷}\mkern-1mu\anonymous }\;{}\<[13]%
\>[13]{}\mathbin{:}\;\Conid{El}_{\text{W}}\;\Varid{a}\;\Varid{→}\;\Varid{∞}\;(\Conid{El}_{\text{P}}\;(\constructor{stream}\;\Varid{a}))\;\Varid{→}\;\Conid{El}_{\text{W}}\;(\constructor{stream}\;\Varid{a}){}\<[E]%
\\
\>[5]{}\hsindent{2}{}\<[7]%
\>[7]{}\anonymous ,\anonymous \;{}\<[13]%
\>[13]{}\mathbin{:}\;\Conid{El}_{\text{W}}\;\Varid{a}\;\Varid{→}\;\Conid{El}_{\text{W}}\;\Varid{b}\;\Varid{→}\;\Conid{El}_{\text{W}}\;(\Varid{a}\;\Varid{⊗}\;\Varid{b}){}\<[E]%
\\
\>[5]{}\hsindent{2}{}\<[7]%
\>[7]{}{\Varid{⌈}\mkern-2mu\anonymous \mkern-0.5mu\Varid{⌉}}\;{}\<[13]%
\>[13]{}\mathbin{:}\;\Conid{A}\;\Varid{→}\;\Conid{El}_{\text{W}}\;\Varid{⌈}\;\Conid{A}\;\Varid{⌉}{}\<[E]%
\ColumnHook
\end{hscode}\resethooks
Note that only those constructor arguments which are delayed are
represented as programs in the definition of \ensuremath{\Conid{El}_{\text{W}}}---the other
arguments can be viewed as ``strict''.
Note also that, unlike in Section~\ref{sec:programs}, the two types
are defined mutually: the WHNF type is included in the type of
programs using the constructor~\ensuremath{\Varid{↓}}. This makes the program type less
usable (the term \ensuremath{\constructor{fst}\;\Varid{p}\;\constructor{∷}\;\Varid{xs}} is not well-typed, for instance), but
avoids some code duplication.
An alternative would be to merge the definitions of \ensuremath{\Conid{El}_{\text{P}}} and \ensuremath{\Conid{El}_{\text{W}}},
and use an additional index to specify which programs are in weak head
normal form.

The type of \ensuremath{\constructor{lab}} may seem a bit strange: the inner and outer streams
are represented differently. One reason for this design choice can be
seen in the definition of \ensuremath{\Varid{lab}_{\text{W}}}:
\begin{hscode}\SaveRestoreHook
\column{B}{@{}>{\hspre}l<{\hspost}@{}}%
\column{3}{@{}>{\hspre}l<{\hspost}@{}}%
\column{5}{@{}>{\hspre}l<{\hspost}@{}}%
\column{22}{@{}>{\hspre}l<{\hspost}@{}}%
\column{42}{@{}>{\hspre}l<{\hspost}@{}}%
\column{79}{@{}>{\hspre}l<{\hspost}@{}}%
\column{E}{@{}>{\hspre}l<{\hspost}@{}}%
\>[3]{}\Varid{lab}_{\text{W}}\;\mathbin{:}\;\Conid{Tree}\;\Conid{A}\;\Varid{→}\;\Conid{El}_{\text{W}}\;(\constructor{stream}\;\Varid{⌈}\;\Conid{Stream}\;\Conid{B}\;\Varid{⌉})\;\Varid{→}\;\Conid{El}_{\text{W}}\;(\constructor{tree}\;\Varid{⌈}\;\Conid{B}\;\Varid{⌉}\;\Varid{⊗}\;\constructor{stream}\;\Varid{⌈}\;\Conid{Stream}\;\Conid{B}\;\Varid{⌉}){}\<[E]%
\\
\>[3]{}\Varid{lab}_{\text{W}}\;\constructor{leaf}\;{}\<[22]%
\>[22]{}\Varid{bss}\;{}\<[42]%
\>[42]{}\mathrel{=}\;(\constructor{leaf},{}\<[79]%
\>[79]{}\Varid{bss}){}\<[E]%
\\
\>[3]{}\Varid{lab}_{\text{W}}\;(\constructor{node}\;\Varid{l}\;\anonymous \;\Varid{r})\;{}\<[22]%
\>[22]{}(\Varid{⌈}\;\Varid{b}\;\constructor{∷}\;\Varid{bs}\;\Varid{⌉}\;\constructor{∷}\;\Varid{bss})\;{}\<[42]%
\>[42]{}\mathrel{=}\;(\constructor{node}\;(\raisebox{0.8ex}[2ex]{\footnotesize ♯}\;\constructor{fst}\;\Varid{x})\;\Varid{⌈}\;\Varid{b}\;\Varid{⌉}\;(\raisebox{0.8ex}[2ex]{\footnotesize ♯}\;\constructor{fst}\;\Varid{y}),{}\<[79]%
\>[79]{}\Varid{⌈}\;\raisebox{0.8ex}[2ex]{\footnotesize ♭}\;\Varid{bs}\;\Varid{⌉}\;\constructor{∷}\;\raisebox{0.8ex}[2ex]{\footnotesize ♯}\;\constructor{snd}\;\Varid{y}){}\<[E]%
\\
\>[3]{}\hsindent{2}{}\<[5]%
\>[5]{}\mathbf{where}{}\<[E]%
\\
\>[3]{}\hsindent{2}{}\<[5]%
\>[5]{}\Varid{x}\;\mathrel{=}\;\constructor{lab}\;(\raisebox{0.8ex}[2ex]{\footnotesize ♭}\;\Varid{l})\;(\raisebox{0.8ex}[2ex]{\footnotesize ♭}\;\Varid{bss}){}\<[E]%
\\
\>[3]{}\hsindent{2}{}\<[5]%
\>[5]{}\Varid{y}\;\mathrel{=}\;\constructor{lab}\;(\raisebox{0.8ex}[2ex]{\footnotesize ♭}\;\Varid{r})\;(\constructor{snd}\;\Varid{x}){}\<[E]%
\ColumnHook
\end{hscode}\resethooks
Consider the second clause. If \ensuremath{\Varid{lab}_{\text{W}}} had the type
\begin{hscode}\SaveRestoreHook
\column{B}{@{}>{\hspre}l<{\hspost}@{}}%
\column{3}{@{}>{\hspre}l<{\hspost}@{}}%
\column{E}{@{}>{\hspre}l<{\hspost}@{}}%
\>[3]{}\Conid{Tree}\;\Conid{A}\;\Varid{→}\;\Conid{El}_{\text{W}}\;(\constructor{stream}\;(\constructor{stream}\;\Varid{b}))\;\Varid{→}\;\Conid{El}_{\text{W}}\;(\constructor{tree}\;\Varid{b}\;\Varid{⊗}\;\constructor{stream}\;(\constructor{stream}\;\Varid{b})),{}\<[E]%
\ColumnHook
\end{hscode}\resethooks
then the analogue of \ensuremath{\Varid{bs}} would be a \emph{program}, but the head of
the resulting stream of streams (\ensuremath{\Varid{⌈}\;\raisebox{0.8ex}[2ex]{\footnotesize ♭}\;\Varid{bs}\;\Varid{⌉}} in the definition above)
must be a WHNF.
The use of ``raw'' inner streams also means that the input to the
\ensuremath{\Varid{label}} function does not need to be converted.

Note that \ensuremath{\Varid{lab}_{\text{W}}} is non-recursive.
The remainder of \ensuremath{\Varid{whnf}} is straightforward to implement using
structural recursion:\\
\begin{minipage}{0.5\linewidth}
  \begin{hscode}\SaveRestoreHook
\column{B}{@{}>{\hspre}l<{\hspost}@{}}%
\column{3}{@{}>{\hspre}l<{\hspost}@{}}%
\column{E}{@{}>{\hspre}l<{\hspost}@{}}%
\>[3]{}\Varid{fst}_{\text{W}}\;\mathbin{:}\;\Conid{El}_{\text{W}}\;(\Varid{a}\;\Varid{⊗}\;\Varid{b})\;\Varid{→}\;\Conid{El}_{\text{W}}\;\Varid{a}{}\<[E]%
\\
\>[3]{}\Varid{fst}_{\text{W}}\;(\Varid{x},\Varid{y})\;\mathrel{=}\;\Varid{x}{}\<[E]%
\\[\blanklineskip]%
\>[3]{}\Varid{snd}_{\text{W}}\;\mathbin{:}\;\Conid{El}_{\text{W}}\;(\Varid{a}\;\Varid{⊗}\;\Varid{b})\;\Varid{→}\;\Conid{El}_{\text{W}}\;\Varid{b}{}\<[E]%
\\
\>[3]{}\Varid{snd}_{\text{W}}\;(\Varid{x},\Varid{y})\;\mathrel{=}\;\Varid{y}{}\<[E]%
\ColumnHook
\end{hscode}\resethooks
\end{minipage}
\begin{minipage}{0.49\linewidth}
  \begin{hscode}\SaveRestoreHook
\column{B}{@{}>{\hspre}l<{\hspost}@{}}%
\column{3}{@{}>{\hspre}l<{\hspost}@{}}%
\column{22}{@{}>{\hspre}l<{\hspost}@{}}%
\column{E}{@{}>{\hspre}l<{\hspost}@{}}%
\>[3]{}\Varid{whnf}\;\mathbin{:}\;\Conid{El}_{\text{P}}\;\Varid{a}\;\Varid{→}\;\Conid{El}_{\text{W}}\;\Varid{a}{}\<[E]%
\\
\>[3]{}\Varid{whnf}\;(\Varid{↓}\;\Varid{w})\;{}\<[22]%
\>[22]{}\mathrel{=}\;\Varid{w}{}\<[E]%
\\
\>[3]{}\Varid{whnf}\;(\constructor{fst}\;\Varid{p})\;{}\<[22]%
\>[22]{}\mathrel{=}\;\Varid{fst}_{\text{W}}\;(\Varid{whnf}\;\Varid{p}){}\<[E]%
\\
\>[3]{}\Varid{whnf}\;(\constructor{snd}\;\Varid{p})\;{}\<[22]%
\>[22]{}\mathrel{=}\;\Varid{snd}_{\text{W}}\;(\Varid{whnf}\;\Varid{p}){}\<[E]%
\\
\>[3]{}\Varid{whnf}\;(\constructor{lab}\;\Varid{t}\;\Varid{bss})\;{}\<[22]%
\>[22]{}\mathrel{=}\;\Varid{lab}_{\text{W}}\;\Varid{t}\;(\Varid{whnf}\;\Varid{bss}){}\<[E]%
\ColumnHook
\end{hscode}\resethooks
\end{minipage}
It is also easy to define \ensuremath{\Varid{⟦}\mkern-1mu\anonymous \mkern1mu\Varid{⟧}_{\text{W}}} and \ensuremath{\Varid{⟦}\mkern-1mu\anonymous \mkern1mu\Varid{⟧}_{\text{P}}}. These definitions use a
lexicographic combination of guarded corecursion and structural
recursion (see Section~\ref{sec:coinduction}):
\begin{hscode}\SaveRestoreHook
\column{B}{@{}>{\hspre}l<{\hspost}@{}}%
\column{3}{@{}>{\hspre}l<{\hspost}@{}}%
\column{20}{@{}>{\hspre}l<{\hspost}@{}}%
\column{E}{@{}>{\hspre}l<{\hspost}@{}}%
\>[B]{}\mathbf{mutual}{}\<[E]%
\\
\>[B]{}\hsindent{3}{}\<[3]%
\>[3]{}\Varid{⟦}\mkern-1mu\anonymous \mkern1mu\Varid{⟧}_{\text{W}}\;\mathbin{:}\;\Conid{El}_{\text{W}}\;\Varid{a}\;\Varid{→}\;\Conid{El}\;\Varid{a}{}\<[E]%
\\
\>[B]{}\hsindent{3}{}\<[3]%
\>[3]{}\Varid{⟦}\;\constructor{leaf}\;\Varid{⟧}_{\text{W}}\;{}\<[20]%
\>[20]{}\mathrel{=}\;\constructor{leaf}{}\<[E]%
\\
\>[B]{}\hsindent{3}{}\<[3]%
\>[3]{}\Varid{⟦}\;\constructor{node}\;\Varid{l}\;\Varid{x}\;\Varid{r}\;\Varid{⟧}_{\text{W}}\;{}\<[20]%
\>[20]{}\mathrel{=}\;\constructor{node}\;(\raisebox{0.8ex}[2ex]{\footnotesize ♯}\;\Varid{⟦}\;\raisebox{0.8ex}[2ex]{\footnotesize ♭}\;\Varid{l}\;\Varid{⟧}_{\text{P}})\;\Varid{⟦}\;\Varid{x}\;\Varid{⟧}_{\text{W}}\;(\raisebox{0.8ex}[2ex]{\footnotesize ♯}\;\Varid{⟦}\;\raisebox{0.8ex}[2ex]{\footnotesize ♭}\;\Varid{r}\;\Varid{⟧}_{\text{P}}){}\<[E]%
\\
\>[B]{}\hsindent{3}{}\<[3]%
\>[3]{}\Varid{⟦}\;\Varid{x}\;\constructor{∷}\;\Varid{xs}\;\Varid{⟧}_{\text{W}}\;{}\<[20]%
\>[20]{}\mathrel{=}\;\Varid{⟦}\;\Varid{x}\;\Varid{⟧}_{\text{W}}\;\constructor{∷}\;\raisebox{0.8ex}[2ex]{\footnotesize ♯}\;\Varid{⟦}\;\raisebox{0.8ex}[2ex]{\footnotesize ♭}\;\Varid{xs}\;\Varid{⟧}_{\text{P}}{}\<[E]%
\\
\>[B]{}\hsindent{3}{}\<[3]%
\>[3]{}\Varid{⟦}\;(\Varid{x},\Varid{y})\;\Varid{⟧}_{\text{W}}\;{}\<[20]%
\>[20]{}\mathrel{=}\;(\Varid{⟦}\;\Varid{x}\;\Varid{⟧}_{\text{W}},\Varid{⟦}\;\Varid{y}\;\Varid{⟧}_{\text{W}}){}\<[E]%
\\
\>[B]{}\hsindent{3}{}\<[3]%
\>[3]{}\Varid{⟦}\;\Varid{⌈}\;\Varid{x}\;\Varid{⌉}\;\Varid{⟧}_{\text{W}}\;{}\<[20]%
\>[20]{}\mathrel{=}\;\Varid{x}{}\<[E]%
\\[\blanklineskip]%
\>[B]{}\hsindent{3}{}\<[3]%
\>[3]{}\Varid{⟦}\mkern-1mu\anonymous \mkern1mu\Varid{⟧}_{\text{P}}\;\mathbin{:}\;\Conid{El}_{\text{P}}\;\Varid{a}\;\Varid{→}\;\Conid{El}\;\Varid{a}{}\<[E]%
\\
\>[B]{}\hsindent{3}{}\<[3]%
\>[3]{}\Varid{⟦}\;\Varid{p}\;\Varid{⟧}_{\text{P}}\;\mathrel{=}\;\Varid{⟦}\;\Varid{whnf}\;\Varid{p}\;\Varid{⟧}_{\text{W}}{}\<[E]%
\ColumnHook
\end{hscode}\resethooks

Finally we can define \ensuremath{\Varid{label}}:
\begin{hscode}\SaveRestoreHook
\column{B}{@{}>{\hspre}l<{\hspost}@{}}%
\column{3}{@{}>{\hspre}l<{\hspost}@{}}%
\column{E}{@{}>{\hspre}l<{\hspost}@{}}%
\>[3]{}\Varid{label′}\;\mathbin{:}\;\Conid{Tree}\;\Conid{A}\;\Varid{→}\;\Conid{Stream}\;\Conid{B}\;\Varid{→}\;\Conid{El}_{\text{P}}\;(\constructor{tree}\;\Varid{⌈}\;\Conid{B}\;\Varid{⌉}\;\Varid{⊗}\;\constructor{stream}\;\Varid{⌈}\;\Conid{Stream}\;\Conid{B}\;\Varid{⌉}){}\<[E]%
\\
\>[3]{}\Varid{label′}\;\Varid{t}\;\Varid{bs}\;\mathrel{=}\;\constructor{lab}\;\Varid{t}\;(\Varid{↓}\;(\Varid{⌈}\;\Varid{bs}\;\Varid{⌉}\;\constructor{∷}\;\raisebox{0.8ex}[2ex]{\footnotesize ♯}\;\constructor{snd}\;(\Varid{label′}\;\Varid{t}\;\Varid{bs}))){}\<[E]%
\\[\blanklineskip]%
\>[3]{}\Varid{label}\;\mathbin{:}\;\Conid{Tree}\;\Conid{A}\;\Varid{→}\;\Conid{Stream}\;\Conid{B}\;\Varid{→}\;\Conid{Tree}\;\Conid{B}{}\<[E]%
\\
\>[3]{}\Varid{label}\;\Varid{t}\;\Varid{bs}\;\mathrel{=}\;\Varid{⟦}\;\constructor{fst}\;(\Varid{label′}\;\Varid{t}\;\Varid{bs})\;\Varid{⟧}_{\text{P}}{}\<[E]%
\ColumnHook
\end{hscode}\resethooks
Note that the helper function \ensuremath{\Varid{label′}}, which corresponds to the
cyclic part of the original \ensuremath{\Varid{label}}, is defined using guarded
corecursion.

I have proved that the definition of \ensuremath{\Varid{label}} is correct: the resulting
tree has the same shape as the original one, and a breadth-first
traversal of the resulting tree produces a potentially infinite list
of labels which is a prefix of the stream given to \ensuremath{\Varid{label}}.
To state correctness I extended the universe with support for
potentially infinite lists, and added some programs to the \ensuremath{\Conid{El}_{\text{P}}} type.
For details of the statement and proof, see
\citet{accompanying-code-par-2010}.

\section{Making Proofs Guarded}
\label{sec:proofs}

The language-based approach to guardedness introduced in
Section~\ref{sec:programs} has some problems when applied to programs:
\begin{itemize}
\item The interpretive overhead, compared to a direct implementation,
  can be substantial. For instance, computing the $n$-th element of
  the stream \ensuremath{\Varid{fib}} defined in Section~\ref{sec:programs} requires a
  number of additions which is exponential in $n$, whereas if \ensuremath{\Varid{fib}\;\mathrel{=}\;0\;\constructor{∷}\;\raisebox{0.8ex}[2ex]{\footnotesize ♯}\;\Varid{zip\mkern-2mu{}With}\;{\anonymous \mkern-0.5mu\Varid{+}\mkern-2mu\anonymous \mkern3mu}\;\Varid{fib}\;(1\;\constructor{∷}\;\raisebox{0.8ex}[2ex]{\footnotesize ♯}\;\Varid{fib})} is implemented directly in a
  language which uses call-by-need this computation only requires
  $\mathcal{O}(n)$ additions.
  The reason for this discrepancy is that the interpreter \ensuremath{\Varid{⟦}\mkern-1mu\anonymous \mkern1mu\Varid{⟧}_{\text{P}}} does
  not preserve sharing.
  One could perhaps work around this problem by writing a more
  complicated interpreter, but this seems counterproductive: why spend
  effort writing a new interpreter when one is already provided by the
  host programming language (or the underlying hardware)?
\item Proving properties about the interpreted definitions (for
  instance to establish that they are correct) can be awkward, because
  this amounts to proving properties about the interpreter.
\end{itemize}
However, these problems are usually irrelevant for \emph{proofs}: the
run-time complexity of proofs is rarely important, and any proof of a
property is usually as good as any other.
Hence the approach is likely to be more useful for making proofs
guarded, than for making programs guarded.

This section shows how the technique can be applied to proofs.
\citet{hinze} advocates proving stream identities using a uniqueness
property. One example in his paper is the iterate fusion law:
\begin{hscode}\SaveRestoreHook
\column{B}{@{}>{\hspre}l<{\hspost}@{}}%
\column{3}{@{}>{\hspre}l<{\hspost}@{}}%
\column{13}{@{}>{\hspre}l<{\hspost}@{}}%
\column{E}{@{}>{\hspre}l<{\hspost}@{}}%
\>[3]{}\Varid{fusion}\;\mathbin{:}\;{}\<[13]%
\>[13]{}(\Varid{h}\;\mathbin{:}\;\Conid{A}\;\Varid{→}\;\Conid{B})\;\Varid{→}\;(\Varid{f}_1\;\mathbin{:}\;\Conid{A}\;\Varid{→}\;\Conid{A})\;\Varid{→}\;(\Varid{f}_2\;\mathbin{:}\;\Conid{B}\;\Varid{→}\;\Conid{B})\;\Varid{→}\;{}\<[E]%
\\
\>[13]{}((\Varid{x}\;\mathbin{:}\;\Conid{A})\;\Varid{→}\;\Varid{h}\;(\Varid{f}_1\;\Varid{x})\;\Varid{≡}\;\Varid{f}_2\;(\Varid{h}\;\Varid{x}))\;\Varid{→}\;{}\<[E]%
\\
\>[13]{}(\Varid{x}\;\mathbin{:}\;\Conid{A})\;\Varid{→}\;\Varid{map}\;\Varid{h}\;(\Varid{iterate}\;\Varid{f}_1\;\Varid{x})\;≈\;\Varid{iterate}\;\Varid{f}_2\;(\Varid{h}\;\Varid{x}){}\<[E]%
\ColumnHook
\end{hscode}\resethooks
Here \ensuremath{\Varid{map}} and \ensuremath{\Varid{iterate}} are defined as follows:\\
\begin{minipage}{0.5\linewidth}
  \begin{hscode}\SaveRestoreHook
\column{B}{@{}>{\hspre}l<{\hspost}@{}}%
\column{5}{@{}>{\hspre}l<{\hspost}@{}}%
\column{E}{@{}>{\hspre}l<{\hspost}@{}}%
\>[5]{}\Varid{map}\;\mathbin{:}\;(\Conid{A}\;\Varid{→}\;\Conid{B})\;\Varid{→}\;\Conid{Stream}\;\Conid{A}\;\Varid{→}\;\Conid{Stream}\;\Conid{B}{}\<[E]%
\\
\>[5]{}\Varid{map}\;\Varid{f}\;(\Varid{x}\;\constructor{∷}\;\Varid{xs})\;\mathrel{=}\;\Varid{f}\;\Varid{x}\;\constructor{∷}\;\raisebox{0.8ex}[2ex]{\footnotesize ♯}\;\Varid{map}\;\Varid{f}\;(\raisebox{0.8ex}[2ex]{\footnotesize ♭}\;\Varid{xs}){}\<[E]%
\ColumnHook
\end{hscode}\resethooks
\end{minipage}
\begin{minipage}{0.49\linewidth}
  \begin{hscode}\SaveRestoreHook
\column{B}{@{}>{\hspre}l<{\hspost}@{}}%
\column{5}{@{}>{\hspre}l<{\hspost}@{}}%
\column{E}{@{}>{\hspre}l<{\hspost}@{}}%
\>[5]{}\Varid{iterate}\;\mathbin{:}\;(\Conid{A}\;\Varid{→}\;\Conid{A})\;\Varid{→}\;\Conid{A}\;\Varid{→}\;\Conid{Stream}\;\Conid{A}{}\<[E]%
\\
\>[5]{}\Varid{iterate}\;\Varid{f}\;\Varid{x}\;\mathrel{=}\;\Varid{x}\;\constructor{∷}\;\raisebox{0.8ex}[2ex]{\footnotesize ♯}\;\Varid{iterate}\;\Varid{f}\;(\Varid{f}\;\Varid{x}){}\<[E]%
\ColumnHook
\end{hscode}\resethooks
\end{minipage}
Hinze proves the iterate fusion law by establishing that the left and
right hand sides both satisfy the same guarded equation, \ensuremath{\Varid{f}\;\Varid{x}\;≈\;\Varid{h}\;\Varid{x}\;\constructor{∷}\;\raisebox{0.8ex}[2ex]{\footnotesize ♯}\;\Varid{f}\;(\Varid{f}_1\;\Varid{x})} (where \ensuremath{\Varid{f}} is the ``unknown variable''):
\begin{hscode}\SaveRestoreHook
\column{B}{@{}>{\hspre}l<{\hspost}@{}}%
\column{3}{@{}>{\hspre}l<{\hspost}@{}}%
\column{38}{@{}>{\hspre}l<{\hspost}@{}}%
\column{E}{@{}>{\hspre}l<{\hspost}@{}}%
\>[3]{}\Varid{map}\;\Varid{h}\;(\Varid{iterate}\;\Varid{f}_1\;\Varid{x})\;{}\<[38]%
\>[38]{}{\ ≈\mkern-6mu⟨}\;\text{by definition}{\mkern4mu⟩}{}\<[E]%
\\
\>[3]{}\Varid{h}\;\Varid{x}\;\constructor{∷}\;\raisebox{0.8ex}[2ex]{\footnotesize ♯}\;\Varid{map}\;\Varid{h}\;(\Varid{iterate}\;\Varid{f}_1\;(\Varid{f}_1\;\Varid{x}))\;{}\<[38]%
\>[38]{} {}\<[E]%
\\
\>[3]{}\vspace{1ex}{}\<[E]%
\\
\>[3]{}\Varid{h}\;\Varid{x}\;\constructor{∷}\;\raisebox{0.8ex}[2ex]{\footnotesize ♯}\;\Varid{iterate}\;\Varid{f}_2\;(\Varid{h}\;(\Varid{f}_1\;\Varid{x}))\;{}\<[38]%
\>[38]{}{\ ≈\mkern-6mu⟨}\;\text{assumption}{\mkern4mu⟩}{}\<[E]%
\\
\>[3]{}\Varid{h}\;\Varid{x}\;\constructor{∷}\;\raisebox{0.8ex}[2ex]{\footnotesize ♯}\;\Varid{iterate}\;\Varid{f}_2\;(\Varid{f}_2\;(\Varid{h}\;\Varid{x}))\;{}\<[38]%
\>[38]{}{\ ≈\mkern-6mu⟨}\;\text{by definition}{\mkern4mu⟩}{}\<[E]%
\\
\>[3]{}\Varid{iterate}\;\Varid{f}_2\;(\Varid{h}\;\Varid{x}){}\<[E]%
\ColumnHook
\end{hscode}\resethooks
The separately proved\footnote{Hinze proves this using a method
  described by \citet{rutten}, which in fact is closely related to the
  method described here, see Section~\ref{sec:related-work}.} fact
that the equation has a unique solution then implies that \ensuremath{\Varid{map}\;\Varid{h}\;(\Varid{iterate}\;\Varid{f}_1\;\Varid{x})} and \ensuremath{\Varid{iterate}\;\Varid{f}_2\;(\Varid{h}\;\Varid{x})} are equal.

Note that the proof above is almost a proof by guarded coinduction:
the two equational reasoning blocks can be joined by an application of
the coinductive hypothesis. However, the second block uses
transitivity, thus destroying guardedness.
We can work around this problem by following the approach introduced
in Section~\ref{sec:programs}.
Let us define a language of equality proof ``programs'' as follows:
\begin{hscode}\SaveRestoreHook
\column{B}{@{}>{\hspre}l<{\hspost}@{}}%
\column{3}{@{}>{\hspre}l<{\hspost}@{}}%
\column{5}{@{}>{\hspre}l<{\hspost}@{}}%
\column{13}{@{}>{\hspre}l<{\hspost}@{}}%
\column{E}{@{}>{\hspre}l<{\hspost}@{}}%
\>[3]{}\mathbf{data}\;{\anonymous \mkern-4mu≈_{\text{P}}\mkern-6mu\anonymous }\;\mathbin{:}\;\Conid{Stream}\;\Conid{A}\;\Varid{→}\;\Conid{Stream}\;\Conid{A}\;\Varid{→}\;\Conid{Set}\;\mathbf{where}{}\<[E]%
\\
\>[3]{}\hsindent{2}{}\<[5]%
\>[5]{}{\anonymous \mkern1mu\constructor{∷}\mkern-1mu\anonymous }\;{}\<[13]%
\>[13]{}\mathbin{:}\;(\Varid{x}\;\mathbin{:}\;\Conid{A})\;\Varid{→}\;\Varid{∞}\;(\raisebox{0.8ex}[2ex]{\footnotesize ♭}\;\Varid{xs}\;≈_{\text{P}}\;\raisebox{0.8ex}[2ex]{\footnotesize ♭}\;\Varid{ys})\;\Varid{→}\;\Varid{x}\;\constructor{∷}\;\Varid{xs}\;≈_{\text{P}}\;\Varid{x}\;\constructor{∷}\;\Varid{ys}{}\<[E]%
\\
\>[3]{}\hsindent{2}{}\<[5]%
\>[5]{}{\anonymous \mkern-6mu{\ ≈\mkern-6mu⟨}\mkern-1.3mu\anonymous \mkern0.5mu⟩\mkern-2mu\anonymous }\;{}\<[13]%
\>[13]{}\mathbin{:}\;(\Varid{xs}\;\mathbin{:}\;\Conid{Stream}\;\Conid{A})\;\Varid{→}\;\Varid{xs}\;≈_{\text{P}}\;\Varid{ys}\;\Varid{→}\;\Varid{ys}\;≈_{\text{P}}\;\Varid{zs}\;\Varid{→}\;\Varid{xs}\;≈_{\text{P}}\;\Varid{zs}{}\<[E]%
\\
\>[3]{}\hsindent{2}{}\<[5]%
\>[5]{}{\anonymous \mkern-4mu\ \square}\;{}\<[13]%
\>[13]{}\mathbin{:}\;(\Varid{xs}\;\mathbin{:}\;\Conid{Stream}\;\Conid{A})\;\Varid{→}\;\Varid{xs}\;≈_{\text{P}}\;\Varid{xs}{}\<[E]%
\ColumnHook
\end{hscode}\resethooks
The last two constructors represent transitivity and reflexivity,
respectively.
Note that the transitivity constructor is inductive; a coinductive
transitivity constructor would make the relation trivial (see
\citet{danielsson-altenkirch-subtyping}).
The somewhat odd names were chosen to make the proof of the iterate
fusion law more readable, following \citet{norell}. Just remember that
\ensuremath{{\anonymous \mkern-6mu{\ ≈\mkern-6mu⟨}\mkern-1.3mu\anonymous \mkern0.5mu⟩\mkern-2mu\anonymous }} and \ensuremath{{\anonymous \mkern-4mu\ \square}} are both weakly binding, with \ensuremath{{\anonymous \mkern-6mu{\ ≈\mkern-6mu⟨}\mkern-1.3mu\anonymous \mkern0.5mu⟩\mkern-2mu\anonymous }} right
associative and binding weaker than \ensuremath{{\anonymous \mkern-4mu\ \square}}:
\begin{hscode}\SaveRestoreHook
\column{B}{@{}>{\hspre}l<{\hspost}@{}}%
\column{3}{@{}>{\hspre}l<{\hspost}@{}}%
\column{5}{@{}>{\hspre}l<{\hspost}@{}}%
\column{13}{@{}>{\hspre}l<{\hspost}@{}}%
\column{40}{@{}>{\hspre}l<{\hspost}@{}}%
\column{E}{@{}>{\hspre}l<{\hspost}@{}}%
\>[3]{}\Varid{fusion}\;\mathbin{:}\;{}\<[13]%
\>[13]{}(\Varid{h}\;\mathbin{:}\;\Conid{A}\;\Varid{→}\;\Conid{B})\;\Varid{→}\;(\Varid{f}_1\;\mathbin{:}\;\Conid{A}\;\Varid{→}\;\Conid{A})\;\Varid{→}\;(\Varid{f}_2\;\mathbin{:}\;\Conid{B}\;\Varid{→}\;\Conid{B})\;\Varid{→}\;{}\<[E]%
\\
\>[13]{}((\Varid{x}\;\mathbin{:}\;\Conid{A})\;\Varid{→}\;\Varid{h}\;(\Varid{f}_1\;\Varid{x})\;\Varid{≡}\;\Varid{f}_2\;(\Varid{h}\;\Varid{x}))\;\Varid{→}\;{}\<[E]%
\\
\>[13]{}(\Varid{x}\;\mathbin{:}\;\Conid{A})\;\Varid{→}\;\Varid{map}\;\Varid{h}\;(\Varid{iterate}\;\Varid{f}_1\;\Varid{x})\;≈_{\text{P}}\;\Varid{iterate}\;\Varid{f}_2\;(\Varid{h}\;\Varid{x}){}\<[E]%
\\
\>[3]{}\Varid{fusion}\;\Varid{h}\;\Varid{f}_1\;\Varid{f}_2\;\Varid{hyp}\;\Varid{x}\;\mathrel{=}\;{}\<[E]%
\\
\>[3]{}\hsindent{2}{}\<[5]%
\>[5]{}\Varid{map}\;\Varid{h}\;(\Varid{iterate}\;\Varid{f}_1\;\Varid{x})\;{}\<[40]%
\>[40]{}{\ ≈\mkern-6mu⟨}\;\Varid{by}\;\Varid{definition}\;⟩\;{}\<[E]%
\\
\>[3]{}\hsindent{2}{}\<[5]%
\>[5]{}\Varid{h}\;\Varid{x}\;\constructor{∷}\;\raisebox{0.8ex}[2ex]{\footnotesize ♯}\;\Varid{map}\;\Varid{h}\;(\Varid{iterate}\;\Varid{f}_1\;(\Varid{f}_1\;\Varid{x}))\;{}\<[40]%
\>[40]{}{\ ≈\mkern-6mu⟨}\;\Varid{h}\;\Varid{x}\;\constructor{∷}\;\raisebox{0.8ex}[2ex]{\footnotesize ♯}\;\Varid{fusion}\;\Varid{h}\;\Varid{f}_1\;\Varid{f}_2\;\Varid{hyp}\;(\Varid{f}_1\;\Varid{x})\;⟩\;{}\<[E]%
\\
\>[3]{}\hsindent{2}{}\<[5]%
\>[5]{}\Varid{h}\;\Varid{x}\;\constructor{∷}\;\raisebox{0.8ex}[2ex]{\footnotesize ♯}\;\Varid{iterate}\;\Varid{f}_2\;(\Varid{h}\;(\Varid{f}_1\;\Varid{x}))\;{}\<[40]%
\>[40]{}{\ ≈\mkern-6mu⟨}\;\Varid{h}\;\Varid{x}\;\constructor{∷}\;\raisebox{0.8ex}[2ex]{\footnotesize ♯}\;\textit{iterate-cong}\;\Varid{f}_2\;(\Varid{hyp}\;\Varid{x})\;⟩\;{}\<[E]%
\\
\>[3]{}\hsindent{2}{}\<[5]%
\>[5]{}\Varid{h}\;\Varid{x}\;\constructor{∷}\;\raisebox{0.8ex}[2ex]{\footnotesize ♯}\;\Varid{iterate}\;\Varid{f}_2\;(\Varid{f}_2\;(\Varid{h}\;\Varid{x}))\;{}\<[40]%
\>[40]{}{\ ≈\mkern-6mu⟨}\;\Varid{by}\;\Varid{definition}\;⟩\;{}\<[E]%
\\
\>[3]{}\hsindent{2}{}\<[5]%
\>[5]{}\Varid{iterate}\;\Varid{f}_2\;(\Varid{h}\;\Varid{x})\;{}\<[40]%
\>[40]{}\ \square{}\<[E]%
\ColumnHook
\end{hscode}\resethooks
Note that the definition of \ensuremath{\Varid{fusion}} is guarded. The definition uses
some simple lemmas (\ensuremath{\textit{iterate-cong}}, \ensuremath{\Varid{by}} and \ensuremath{\Varid{definition}}), which are
omitted here.

In order to finish the proof of the iterate fusion law we have to show
that \ensuremath{{\anonymous \mkern-4mu≈_{\text{P}}\mkern-6mu\anonymous }} is sound with respect to \ensuremath{{\anonymous \mkern-4mu≈\mkern-6mu\anonymous }}.
To do this one can first define a type of WHNFs:
\begin{hscode}\SaveRestoreHook
\column{B}{@{}>{\hspre}l<{\hspost}@{}}%
\column{3}{@{}>{\hspre}l<{\hspost}@{}}%
\column{5}{@{}>{\hspre}l<{\hspost}@{}}%
\column{E}{@{}>{\hspre}l<{\hspost}@{}}%
\>[3]{}\mathbf{data}\;{\anonymous \mkern-4mu≈_{\text{W}}\mkern-6mu\anonymous }\;\mathbin{:}\;\Conid{Stream}\;\Conid{A}\;\Varid{→}\;\Conid{Stream}\;\Conid{A}\;\Varid{→}\;\Conid{Set}\;\mathbf{where}{}\<[E]%
\\
\>[3]{}\hsindent{2}{}\<[5]%
\>[5]{}{\anonymous \mkern1mu\constructor{∷}\mkern-1mu\anonymous }\;\mathbin{:}\;(\Varid{x}\;\mathbin{:}\;\Conid{A})\;\Varid{→}\;\raisebox{0.8ex}[2ex]{\footnotesize ♭}\;\Varid{xs}\;≈_{\text{P}}\;\raisebox{0.8ex}[2ex]{\footnotesize ♭}\;\Varid{ys}\;\Varid{→}\;\Varid{x}\;\constructor{∷}\;\Varid{xs}\;≈_{\text{W}}\;\Varid{x}\;\constructor{∷}\;\Varid{ys}{}\<[E]%
\ColumnHook
\end{hscode}\resethooks
It is easy to establish, by simple case analysis, that this relation
is a preorder:
\begin{hscode}\SaveRestoreHook
\column{B}{@{}>{\hspre}l<{\hspost}@{}}%
\column{3}{@{}>{\hspre}l<{\hspost}@{}}%
\column{11}{@{}>{\hspre}l<{\hspost}@{}}%
\column{E}{@{}>{\hspre}l<{\hspost}@{}}%
\>[3]{}\Varid{refl}_{\text{W}}\;{}\<[11]%
\>[11]{}\mathbin{:}\;(\Varid{xs}\;\mathbin{:}\;\Conid{Stream}\;\Conid{A})\;\Varid{→}\;\Varid{xs}\;≈_{\text{W}}\;\Varid{xs}{}\<[E]%
\\
\>[3]{}\Varid{trans}_{\text{W}}\;{}\<[11]%
\>[11]{}\mathbin{:}\;\Varid{xs}\;≈_{\text{W}}\;\Varid{ys}\;\Varid{→}\;\Varid{ys}\;≈_{\text{W}}\;\Varid{zs}\;\Varid{→}\;\Varid{xs}\;≈_{\text{W}}\;\Varid{zs}{}\<[E]%
\ColumnHook
\end{hscode}\resethooks
It follows by structural recursion that programs can be turned into
WHNFs:
\begin{hscode}\SaveRestoreHook
\column{B}{@{}>{\hspre}l<{\hspost}@{}}%
\column{3}{@{}>{\hspre}l<{\hspost}@{}}%
\column{31}{@{}>{\hspre}l<{\hspost}@{}}%
\column{E}{@{}>{\hspre}l<{\hspost}@{}}%
\>[3]{}\Varid{whnf}\;\mathbin{:}\;\Varid{xs}\;≈_{\text{P}}\;\Varid{ys}\;\Varid{→}\;\Varid{xs}\;≈_{\text{W}}\;\Varid{ys}{}\<[E]%
\\
\>[3]{}\Varid{whnf}\;(\Varid{x}\;\constructor{∷}\;\Varid{xs}\mkern-5mu≈\mkern-6mu\Varid{ys})\;{}\<[31]%
\>[31]{}\mathrel{=}\;\Varid{x}\;\constructor{∷}\;\raisebox{0.8ex}[2ex]{\footnotesize ♭}\;\Varid{xs}\mkern-5mu≈\mkern-6mu\Varid{ys}{}\<[E]%
\\
\>[3]{}\Varid{whnf}\;(\Varid{xs}\;{\ ≈\mkern-6mu⟨}\;\Varid{xs}\mkern-5mu≈\mkern-6mu\Varid{ys}\;⟩\;\Varid{ys}\mkern-5mu≈\mkern-6mu\Varid{zs})\;{}\<[31]%
\>[31]{}\mathrel{=}\;\Varid{trans}_{\text{W}}\;(\Varid{whnf}\;\Varid{xs}\mkern-5mu≈\mkern-6mu\Varid{ys})\;(\Varid{whnf}\;\Varid{ys}\mkern-5mu≈\mkern-6mu\Varid{zs}){}\<[E]%
\\
\>[3]{}\Varid{whnf}\;(\Varid{xs}\;\ \square)\;{}\<[31]%
\>[31]{}\mathrel{=}\;\Varid{refl}_{\text{W}}\;\Varid{xs}{}\<[E]%
\ColumnHook
\end{hscode}\resethooks
Finally soundness can be proved using guarded corecursion:
\begin{hscode}\SaveRestoreHook
\column{B}{@{}>{\hspre}l<{\hspost}@{}}%
\column{3}{@{}>{\hspre}l<{\hspost}@{}}%
\column{5}{@{}>{\hspre}l<{\hspost}@{}}%
\column{E}{@{}>{\hspre}l<{\hspost}@{}}%
\>[3]{}\mathbf{mutual}{}\<[E]%
\\
\>[3]{}\hsindent{2}{}\<[5]%
\>[5]{}\Varid{sound}_{\text{W}}\;\mathbin{:}\;\Varid{xs}\;≈_{\text{W}}\;\Varid{ys}\;\Varid{→}\;\Varid{xs}\;≈\;\Varid{ys}{}\<[E]%
\\
\>[3]{}\hsindent{2}{}\<[5]%
\>[5]{}\Varid{sound}_{\text{W}}\;(\Varid{x}\;\constructor{∷}\;\Varid{xs}\mkern-5mu≈\mkern-6mu\Varid{ys})\;\mathrel{=}\;\Varid{x}\;\constructor{∷}\;\raisebox{0.8ex}[2ex]{\footnotesize ♯}\;\Varid{sound}_{\text{P}}\;\Varid{xs}\mkern-5mu≈\mkern-6mu\Varid{ys}{}\<[E]%
\\[\blanklineskip]%
\>[3]{}\hsindent{2}{}\<[5]%
\>[5]{}\Varid{sound}_{\text{P}}\;\mathbin{:}\;\Varid{xs}\;≈_{\text{P}}\;\Varid{ys}\;\Varid{→}\;\Varid{xs}\;≈\;\Varid{ys}{}\<[E]%
\\
\>[3]{}\hsindent{2}{}\<[5]%
\>[5]{}\Varid{sound}_{\text{P}}\;\Varid{xs}\mkern-5mu≈\mkern-6mu\Varid{ys}\;\mathrel{=}\;\Varid{sound}_{\text{W}}\;(\Varid{whnf}\;\Varid{xs}\mkern-5mu≈\mkern-6mu\Varid{ys}){}\<[E]%
\ColumnHook
\end{hscode}\resethooks
Note that there is no need to prove that the application \ensuremath{\Varid{sound}_{\text{P}}\;(\Varid{fusion}\;\Varid{h}\;\Varid{f}_1\;\Varid{f}_2\;\Varid{hyp}\;\Varid{x})} satisfies its intended defining equation,
whatever that would be, or that this equation has a unique solution.

Using the language-based approach to guardedness I have formalised a
number of examples from Hinze's paper, see
\citet{accompanying-code-par-2010}. Rephrasing the proofs using
guarded coinduction turned out to be unproblematic.

As a further example, let us show that the defining equation for \ensuremath{\Varid{fib}}
(see Section~\ref{sec:programs}) has a unique solution. We can state
the problem as follows:
\begin{hscode}\SaveRestoreHook
\column{B}{@{}>{\hspre}l<{\hspost}@{}}%
\column{E}{@{}>{\hspre}l<{\hspost}@{}}%
\>[B]{}\textit{fib-rhs}\;\mathbin{:}\;\Conid{Stream}\;\Conid{ℕ}\;\Varid{→}\;\Conid{Stream}\;\Conid{ℕ}{}\<[E]%
\\
\>[B]{}\textit{fib-rhs}\;\Varid{ns}\;\mathrel{=}\;0\;\constructor{∷}\;\raisebox{0.8ex}[2ex]{\footnotesize ♯}\;\Varid{zip\mkern-2mu{}With}\;{\anonymous \mkern-0.5mu\Varid{+}\mkern-2mu\anonymous \mkern3mu}\;\Varid{ns}\;(1\;\constructor{∷}\;\raisebox{0.8ex}[2ex]{\footnotesize ♯}\;\Varid{ns}){}\<[E]%
\\[\blanklineskip]%
\>[B]{}\textit{fib-unique}\;\mathbin{:}\;(\Varid{ms}\;\Varid{ns}\;\mathbin{:}\;\Conid{Stream}\;\Conid{ℕ})\;\Varid{→}\;\Varid{ms}\;≈\;\textit{fib-rhs}\;\Varid{ms}\;\Varid{→}\;\Varid{ns}\;≈\;\textit{fib-rhs}\;\Varid{ns}\;\Varid{→}\;\Varid{ms}\;≈_{\text{P}}\;\Varid{ns}{}\<[E]%
\ColumnHook
\end{hscode}\resethooks
The type \ensuremath{{\anonymous \mkern-4mu≈_{\text{P}}\mkern-6mu\anonymous }} used here is different from the one used above: the
proof will make use of the congruence of \ensuremath{\Varid{zip\mkern-2mu{}With}}, and the
coinductive hypothesis will be an argument to this congruence, so a
constructor for the congruence is included among the equality proof
programs:
\begin{hscode}\SaveRestoreHook
\column{B}{@{}>{\hspre}l<{\hspost}@{}}%
\column{3}{@{}>{\hspre}l<{\hspost}@{}}%
\column{5}{@{}>{\hspre}l<{\hspost}@{}}%
\column{21}{@{}>{\hspre}l<{\hspost}@{}}%
\column{E}{@{}>{\hspre}l<{\hspost}@{}}%
\>[3]{}\mathbf{data}\;{\anonymous \mkern-4mu≈_{\text{P}}\mkern-6mu\anonymous }\;\mathbin{:}\;\Conid{Stream}\;\Conid{A}\;\Varid{→}\;\Conid{Stream}\;\Conid{A}\;\Varid{→}\;\Conid{Set}\;\mathbf{where}{}\<[E]%
\\
\>[3]{}\hsindent{2}{}\<[5]%
\>[5]{}\ldots{}\<[E]%
\\
\>[3]{}\hsindent{2}{}\<[5]%
\>[5]{}\constructor{zipWith}\textit{-}\constructor{cong}\;\mathbin{:}\;{}\<[21]%
\>[21]{}(\Varid{f}\;\mathbin{:}\;\Conid{A}\;\Varid{→}\;\Conid{A}\;\Varid{→}\;\Conid{A})\;\Varid{→}\;\Varid{xs}_{\text{1}}\;≈_{\text{P}}\;\Varid{ys}_{\text{1}}\;\Varid{→}\;\Varid{xs}_{\text{2}}\;≈_{\text{P}}\;\Varid{ys}_{\text{2}}\;\Varid{→}\;{}\<[E]%
\\
\>[21]{}\Varid{zip\mkern-2mu{}With}\;\Varid{f}\;\Varid{xs}_{\text{1}}\;\Varid{xs}_{\text{2}}\;≈_{\text{P}}\;\Varid{zip\mkern-2mu{}With}\;\Varid{f}\;\Varid{ys}_{\text{1}}\;\Varid{ys}_{\text{2}}{}\<[E]%
\ColumnHook
\end{hscode}\resethooks
It is easy to extend the definition of \ensuremath{\Varid{whnf}} to support
\ensuremath{\constructor{zipWith}\textit{-}\constructor{cong}}, using which we can define \ensuremath{\textit{fib-unique}} as follows:
\begin{hscode}\SaveRestoreHook
\column{B}{@{}>{\hspre}l<{\hspost}@{}}%
\column{3}{@{}>{\hspre}l<{\hspost}@{}}%
\column{15}{@{}>{\hspre}l<{\hspost}@{}}%
\column{42}{@{}>{\hspre}l<{\hspost}@{}}%
\column{E}{@{}>{\hspre}l<{\hspost}@{}}%
\>[B]{}\textit{fib-unique}\;\Varid{ms}\;\Varid{ns}\;\Varid{hyp}_{\text{1}}\;\Varid{hyp}_{\text{2}}\;\mathrel{=}\;{}\<[E]%
\\
\>[B]{}\hsindent{3}{}\<[3]%
\>[3]{}\Varid{ms}\;{}\<[15]%
\>[15]{}{\ ≈\mkern-6mu⟨}\;\Varid{complete}_{\text{P}}\;\Varid{hyp}_{\text{1}}\;⟩\;{}\<[E]%
\\
\>[B]{}\hsindent{3}{}\<[3]%
\>[3]{}\textit{fib-rhs}\;\Varid{ms}\;{}\<[15]%
\>[15]{}{\ ≈\mkern-6mu⟨}\;0\;\constructor{∷}\;\raisebox{0.8ex}[2ex]{\footnotesize ♯}\;\constructor{zipWith}\textit{-}\constructor{cong}\;{\anonymous \mkern-0.5mu\Varid{+}\mkern-2mu\anonymous \mkern3mu}\;{}\<[42]%
\>[42]{}(\textit{fib-unique}\;\Varid{ms}\;\Varid{ns}\;\Varid{hyp}_{\text{1}}\;\Varid{hyp}_{\text{2}})\;{}\<[E]%
\\
\>[42]{}(1\;\constructor{∷}\;\raisebox{0.8ex}[2ex]{\footnotesize ♯}\;\textit{fib-unique}\;\Varid{ms}\;\Varid{ns}\;\Varid{hyp}_{\text{1}}\;\Varid{hyp}_{\text{2}})\;⟩\;{}\<[E]%
\\
\>[B]{}\hsindent{3}{}\<[3]%
\>[3]{}\textit{fib-rhs}\;\Varid{ns}\;{}\<[15]%
\>[15]{}{\ ≈\mkern-6mu⟨}\;\Varid{complete}_{\text{P}}\;(\Varid{sym}\;\Varid{hyp}_{\text{2}})\;⟩\;{}\<[E]%
\\
\>[B]{}\hsindent{3}{}\<[3]%
\>[3]{}\Varid{ns}\;{}\<[15]%
\>[15]{}\ \square{}\<[E]%
\ColumnHook
\end{hscode}\resethooks
Here \ensuremath{\Varid{sym}} is the proof of symmetry of \ensuremath{{\anonymous \mkern-4mu≈\mkern-6mu\anonymous }} from
Section~\ref{sec:coinduction}, and \ensuremath{\Varid{complete}_{\text{P}}} shows that \ensuremath{{\anonymous \mkern-4mu≈_{\text{P}}\mkern-6mu\anonymous }} is
complete with respect to \ensuremath{{\anonymous \mkern-4mu≈\mkern-6mu\anonymous }}:
\begin{hscode}\SaveRestoreHook
\column{B}{@{}>{\hspre}l<{\hspost}@{}}%
\column{E}{@{}>{\hspre}l<{\hspost}@{}}%
\>[B]{}\Varid{complete}_{\text{P}}\;\mathbin{:}\;\Varid{xs}\;≈\;\Varid{ys}\;\Varid{→}\;\Varid{xs}\;≈_{\text{P}}\;\Varid{ys}{}\<[E]%
\\
\>[B]{}\Varid{complete}_{\text{P}}\;(\Varid{x}\;\constructor{∷}\;\Varid{xs}\mkern-5mu≈\mkern-6mu\Varid{ys})\;\mathrel{=}\;\Varid{x}\;\constructor{∷}\;\raisebox{0.8ex}[2ex]{\footnotesize ♯}\;\Varid{complete}_{\text{P}}\;(\raisebox{0.8ex}[2ex]{\footnotesize ♭}\;\Varid{xs}\mkern-5mu≈\mkern-6mu\Varid{ys}){}\<[E]%
\ColumnHook
\end{hscode}\resethooks

\section{Destructors}
\label{sec:destructors}

The following, alternative definition of the Fibonacci sequence is not
directly supported by the framework outlined in previous sections:
\begin{hscode}\SaveRestoreHook
\column{B}{@{}>{\hspre}l<{\hspost}@{}}%
\column{3}{@{}>{\hspre}l<{\hspost}@{}}%
\column{E}{@{}>{\hspre}l<{\hspost}@{}}%
\>[3]{}\Varid{fib}\;\mathbin{:}\;\Conid{Stream}\;\Conid{ℕ}{}\<[E]%
\\
\>[3]{}\Varid{fib}\;\mathrel{=}\;0\;\constructor{∷}\;\raisebox{0.8ex}[2ex]{\footnotesize ♯}\;(1\;\constructor{∷}\;\raisebox{0.8ex}[2ex]{\footnotesize ♯}\;\Varid{zip\mkern-2mu{}With}\;{\anonymous \mkern-0.5mu\Varid{+}\mkern-2mu\anonymous \mkern3mu}\;\Varid{fib}\;(\Varid{tail}\;\Varid{fib})){}\<[E]%
\ColumnHook
\end{hscode}\resethooks
The problem is the use of the destructor \ensuremath{\Varid{tail}}. Unrestricted use of
destructors can lead to non-productive ``definitions'', as
demonstrated by \ensuremath{\Varid{bad}} (see Section~\ref{sec:introduction}). However,
destructors can be incorporated by extending the program type with an
index which indicates when they can be used.

Consider the following type of stream programs:
\begin{hscode}\SaveRestoreHook
\column{B}{@{}>{\hspre}l<{\hspost}@{}}%
\column{3}{@{}>{\hspre}l<{\hspost}@{}}%
\column{5}{@{}>{\hspre}l<{\hspost}@{}}%
\column{15}{@{}>{\hspre}l<{\hspost}@{}}%
\column{18}{@{}>{\hspre}l<{\hspost}@{}}%
\column{23}{@{}>{\hspre}l<{\hspost}@{}}%
\column{38}{@{}>{\hspre}l<{\hspost}@{}}%
\column{41}{@{}>{\hspre}l<{\hspost}@{}}%
\column{58}{@{}>{\hspre}l<{\hspost}@{}}%
\column{E}{@{}>{\hspre}l<{\hspost}@{}}%
\>[3]{}\mathbf{data}\;\Conid{Stream}_{\text{P}}\;\mathbin{:}\;\Conid{Bool}\;\Varid{→}\;\Conid{Set}\;\Varid{→}\;\Conid{Set}_{\text{1}}\;\mathbf{where}{}\<[E]%
\\
\>[3]{}\hsindent{2}{}\<[5]%
\>[5]{}[\mskip1.5mu \anonymous \mskip1.5mu]\;{}\<[15]%
\>[15]{}\mathbin{:}\;{}\<[18]%
\>[18]{}\Varid{∞}\;(\Conid{Stream}_{\text{P}}\;\constructor{true}\;\Conid{A})\;{}\<[41]%
\>[41]{}\Varid{→}\;\Conid{Stream}_{\text{P}}\;\constructor{false}\;{}\<[58]%
\>[58]{}\Conid{A}{}\<[E]%
\\
\>[3]{}\hsindent{2}{}\<[5]%
\>[5]{}{\anonymous \mkern1mu\constructor{∷}\mkern-1mu\anonymous }\;{}\<[15]%
\>[15]{}\mathbin{:}\;{}\<[18]%
\>[18]{}\Conid{A}\;\Varid{→}\;{}\<[23]%
\>[23]{}\Conid{Stream}_{\text{P}}\;\constructor{false}\;{}\<[38]%
\>[38]{}\Conid{A}\;{}\<[41]%
\>[41]{}\Varid{→}\;\Conid{Stream}_{\text{P}}\;\constructor{true}\;{}\<[58]%
\>[58]{}\Conid{A}{}\<[E]%
\\
\>[3]{}\hsindent{2}{}\<[5]%
\>[5]{}\constructor{tail}\;{}\<[15]%
\>[15]{}\mathbin{:}\;{}\<[23]%
\>[23]{}\Conid{Stream}_{\text{P}}\;\constructor{true}\;{}\<[38]%
\>[38]{}\Conid{A}\;{}\<[41]%
\>[41]{}\Varid{→}\;\Conid{Stream}_{\text{P}}\;\constructor{false}\;{}\<[58]%
\>[58]{}\Conid{A}{}\<[E]%
\\
\>[3]{}\hsindent{2}{}\<[5]%
\>[5]{}\constructor{forget}\;{}\<[15]%
\>[15]{}\mathbin{:}\;{}\<[23]%
\>[23]{}\Conid{Stream}_{\text{P}}\;\constructor{true}\;{}\<[38]%
\>[38]{}\Conid{A}\;{}\<[41]%
\>[41]{}\Varid{→}\;\Conid{Stream}_{\text{P}}\;\constructor{false}\;{}\<[58]%
\>[58]{}\Conid{A}{}\<[E]%
\\
\>[3]{}\hsindent{2}{}\<[5]%
\>[5]{}\constructor{zip\mkern-0.8mu With}\;{}\<[15]%
\>[15]{}\mathbin{:}\;{}\<[18]%
\>[18]{}(\Conid{A}\;\Varid{→}\;\Conid{B}\;\Varid{→}\;\Conid{C})\;\Varid{→}\;\Conid{Stream}_{\text{P}}\;\Varid{b}\;\Conid{A}\;\Varid{→}\;\Conid{Stream}_{\text{P}}\;\Varid{b}\;\Conid{B}\;\Varid{→}\;\Conid{Stream}_{\text{P}}\;\Varid{b}\;\Conid{C}{}\<[E]%
\ColumnHook
\end{hscode}\resethooks
The type \ensuremath{\Conid{Stream}_{\text{P}}\;\Varid{b}\;\Conid{A}} stands for streams generated in chunks of size
(at least) one, where the first chunk is guaranteed to be non-empty if
the index \ensuremath{\Varid{b}} is \ensuremath{\constructor{true}}.
The constructor \ensuremath{[\mskip1.5mu \anonymous \mskip1.5mu]} marks the end of a chunk.
Note how the indices ensure that a finished chunk is always non-empty,
and that \ensuremath{\constructor{tail}} may only be used to inspect the chunk currently being
constructed.
The constructor \ensuremath{\constructor{forget}} is used to ``forget'' that a chunk is already
finished; \ensuremath{\constructor{forget}} represents the identity function. This constructor
is used in the implementation of \ensuremath{\Varid{fib}_{\text{P}}} (an alternative would be to
give \ensuremath{\constructor{zip\mkern-0.8mu With}} a more general type):
\begin{hscode}\SaveRestoreHook
\column{B}{@{}>{\hspre}l<{\hspost}@{}}%
\column{3}{@{}>{\hspre}l<{\hspost}@{}}%
\column{E}{@{}>{\hspre}l<{\hspost}@{}}%
\>[3]{}\Varid{fib}_{\text{P}}\;\mathbin{:}\;\Conid{Stream}_{\text{P}}\;\constructor{true}\;\Conid{ℕ}{}\<[E]%
\\
\>[3]{}\Varid{fib}_{\text{P}}\;\mathrel{=}\;0\;\constructor{∷}\;[\mskip1.5mu \raisebox{0.8ex}[2ex]{\footnotesize ♯}\;(1\;\constructor{∷}\;\constructor{zip\mkern-0.8mu With}\;{\anonymous \mkern-0.5mu\Varid{+}\mkern-2mu\anonymous \mkern3mu}\;(\constructor{forget}\;\Varid{fib}_{\text{P}})\;(\constructor{tail}\;\Varid{fib}_{\text{P}}))\mskip1.5mu]{}\<[E]%
\ColumnHook
\end{hscode}\resethooks
The implementation of \ensuremath{\Varid{⟦}\mkern-1mu\anonymous \mkern1mu\Varid{⟧}_{\text{P}}} for this language is very similar to that
for the language in Section~\ref{sec:other-chunk-sizes}, so it is
omitted here.
For details of this implementation, the proof of correctness of
\ensuremath{\Varid{fib}_{\text{P}}}, and the proof of uniqueness of solutions of the defining
equation for \ensuremath{\Varid{fib}_{\text{P}}}, see \citet{accompanying-code-par-2010}.

\section{Other Chunk Sizes}
\label{sec:other-chunk-sizes}

The language of the previous section can be generalised to support
other chunk sizes \citep{accompanying-code-par-2010}.
Larger chunk sizes can provide interesting situations. Consider the
following alternative definition of the function \ensuremath{\Varid{map}} from
Section~\ref{sec:proofs}:
\begin{hscode}\SaveRestoreHook
\column{B}{@{}>{\hspre}l<{\hspost}@{}}%
\column{5}{@{}>{\hspre}l<{\hspost}@{}}%
\column{22}{@{}>{\hspre}l<{\hspost}@{}}%
\column{E}{@{}>{\hspre}l<{\hspost}@{}}%
\>[5]{}\Varid{map}_{\text{2}}\mkern1mu\;\mathbin{:}\;(\Conid{A}\;\Varid{→}\;\Conid{B})\;\Varid{→}\;\Conid{Stream}\;\Conid{A}\;\Varid{→}\;\Conid{Stream}\;\Conid{B}{}\<[E]%
\\
\>[5]{}\Varid{map}_{\text{2}}\mkern1mu\;\Varid{f}\;(\Varid{x}\;\constructor{∷}\;\Varid{xs})\;{}\<[22]%
\>[22]{}\mathbf{with}\;\raisebox{0.8ex}[2ex]{\footnotesize ♭}\;\Varid{xs}{}\<[E]%
\\
\>[5]{}\Varid{map}_{\text{2}}\mkern1mu\;\Varid{f}\;(\Varid{x}\;\constructor{∷}\;\Varid{xs})\;{}\<[22]%
\>[22]{}\mid \;\Varid{y}\;\constructor{∷}\;\Varid{ys}\;\mathrel{=}\;\Varid{f}\;\Varid{x}\;\constructor{∷}\;\raisebox{0.8ex}[2ex]{\footnotesize ♯}\;(\Varid{f}\;\Varid{y}\;\constructor{∷}\;\raisebox{0.8ex}[2ex]{\footnotesize ♯}\;\Varid{map}_{\text{2}}\mkern1mu\;\Varid{f}\;(\raisebox{0.8ex}[2ex]{\footnotesize ♭}\;\Varid{ys})){}\<[E]%
\ColumnHook
\end{hscode}\resethooks
One can show that \ensuremath{\Varid{map}} and \ensuremath{\Varid{map}_{\text{2}}\mkern1mu} are extensionally equal:
\begin{hscode}\SaveRestoreHook
\column{B}{@{}>{\hspre}l<{\hspost}@{}}%
\column{5}{@{}>{\hspre}l<{\hspost}@{}}%
\column{E}{@{}>{\hspre}l<{\hspost}@{}}%
\>[5]{}{\Varid{map}\mkern-5mu≈\mkern-6mu\Varid{map}_{\text{2}}\mkern1mu}\;\mathbin{:}\;(\Varid{f}\;\mathbin{:}\;\Conid{A}\;\Varid{→}\;\Conid{B})\;\Varid{→}\;(\Varid{xs}\;\mathbin{:}\;\Conid{Stream}\;\Conid{A})\;\Varid{→}\;\Varid{map}\;\Varid{f}\;\Varid{xs}\;≈\;\Varid{map}_{\text{2}}\mkern1mu\;\Varid{f}\;\Varid{xs}{}\<[E]%
\ColumnHook
\end{hscode}\resethooks
However, assuming that pattern matching is ``strict'', they are not
interchangeable.
The following definition of the stream of natural numbers is
productive, albeit not guarded:
\begin{hscode}\SaveRestoreHook
\column{B}{@{}>{\hspre}l<{\hspost}@{}}%
\column{3}{@{}>{\hspre}l<{\hspost}@{}}%
\column{E}{@{}>{\hspre}l<{\hspost}@{}}%
\>[3]{}\Varid{nats}\;\mathbin{:}\;\Conid{Stream}\;\Conid{ℕ}{}\<[E]%
\\
\>[3]{}\Varid{nats}\;\mathrel{=}\;0\;\constructor{∷}\;\raisebox{0.8ex}[2ex]{\footnotesize ♯}\;\Varid{map}\;\constructor{suc}\;\Varid{nats}{}\<[E]%
\ColumnHook
\end{hscode}\resethooks
The definition that we get by replacing \ensuremath{\Varid{map}} by \ensuremath{\Varid{map}_{\text{2}}}, on the other
hand, is not productive:
\begin{hscode}\SaveRestoreHook
\column{B}{@{}>{\hspre}l<{\hspost}@{}}%
\column{3}{@{}>{\hspre}l<{\hspost}@{}}%
\column{E}{@{}>{\hspre}l<{\hspost}@{}}%
\>[3]{}\Varid{nats}_{\text{2}}\;\mathbin{:}\;\Conid{Stream}\;\Conid{ℕ}{}\<[E]%
\\
\>[3]{}\Varid{nats}_{\text{2}}\;\mathrel{=}\;0\;\constructor{∷}\;\raisebox{0.8ex}[2ex]{\footnotesize ♯}\;\Varid{map}_{\text{2}}\mkern1mu\;\constructor{suc}\;\Varid{nats}_{\text{2}}{}\<[E]%
\ColumnHook
\end{hscode}\resethooks
The first element of \ensuremath{\Varid{nats}_{\text{2}}} is \ensuremath{0}, but \ensuremath{\Varid{map}_{\text{2}}} needs to access the
first \emph{two} elements of its argument stream in order to output
anything.

We can perhaps get a better picture of the situation above using the
following language:
\begin{hscode}\SaveRestoreHook
\column{B}{@{}>{\hspre}l<{\hspost}@{}}%
\column{3}{@{}>{\hspre}l<{\hspost}@{}}%
\column{5}{@{}>{\hspre}l<{\hspost}@{}}%
\column{11}{@{}>{\hspre}l<{\hspost}@{}}%
\column{E}{@{}>{\hspre}l<{\hspost}@{}}%
\>[3]{}\mathbf{data}\;\Conid{Stream}_{\text{P}}\;(\Varid{m}\;\mathbin{:}\;\Conid{ℕ})\;\mathbin{:}\;\Conid{ℕ}\;\Varid{→}\;\Conid{Set}\;\Varid{→}\;\Conid{Set}_{\text{1}}\;\mathbf{where}{}\<[E]%
\\
\>[3]{}\hsindent{2}{}\<[5]%
\>[5]{}[\mskip1.5mu \anonymous \mskip1.5mu]\;{}\<[11]%
\>[11]{}\mathbin{:}\;\Varid{∞}\;(\Conid{Stream}_{\text{P}}\;\Varid{m}\;\Varid{m}\;\Conid{A})\;\Varid{→}\;\Conid{Stream}_{\text{P}}\;\Varid{m}\;0\;\Conid{A}{}\<[E]%
\\
\>[3]{}\hsindent{2}{}\<[5]%
\>[5]{}{\anonymous \mkern1mu\constructor{∷}\mkern-1mu\anonymous }\;{}\<[11]%
\>[11]{}\mathbin{:}\;\Conid{A}\;\Varid{→}\;\Conid{Stream}_{\text{P}}\;\Varid{m}\;\Varid{n}\;\Conid{A}\;\Varid{→}\;\Conid{Stream}_{\text{P}}\;\Varid{m}\;(\constructor{suc}\;\Varid{n})\;\Conid{A}{}\<[E]%
\\
\>[3]{}\hsindent{2}{}\<[5]%
\>[5]{}\constructor{map}\;{}\<[11]%
\>[11]{}\mathbin{:}\;(\Conid{A}\;\Varid{→}\;\Conid{B})\;\Varid{→}\;\Conid{Stream}_{\text{P}}\;\Varid{m}\;\Varid{n}\;\Conid{A}\;\Varid{→}\;\Conid{Stream}_{\text{P}}\;\Varid{m}\;\Varid{n}\;\Conid{B}{}\<[E]%
\ColumnHook
\end{hscode}\resethooks
\ensuremath{\Conid{Stream}_{\text{P}}\;\Varid{m}\;\Varid{n}\;\Conid{A}} is a language of programs which generate streams of
\ensuremath{\Conid{A}}s in chunks of size \ensuremath{\Varid{m}}, where the first chunk has size \ensuremath{\Varid{n}}.
We can define WHNFs and the \ensuremath{\Varid{whnf}} function as follows:
\begin{hscode}\SaveRestoreHook
\column{B}{@{}>{\hspre}l<{\hspost}@{}}%
\column{3}{@{}>{\hspre}l<{\hspost}@{}}%
\column{5}{@{}>{\hspre}l<{\hspost}@{}}%
\column{10}{@{}>{\hspre}l<{\hspost}@{}}%
\column{20}{@{}>{\hspre}l<{\hspost}@{}}%
\column{21}{@{}>{\hspre}l<{\hspost}@{}}%
\column{E}{@{}>{\hspre}l<{\hspost}@{}}%
\>[3]{}\mathbf{data}\;\Conid{Stream}_{\text{W}}\;(\Varid{m}\;\mathbin{:}\;\Conid{ℕ})\;\mathbin{:}\;\Conid{ℕ}\;\Varid{→}\;\Conid{Set}\;\Varid{→}\;\Conid{Set}_{\text{1}}\;\mathbf{where}{}\<[E]%
\\
\>[3]{}\hsindent{2}{}\<[5]%
\>[5]{}[\mskip1.5mu \anonymous \mskip1.5mu]\;{}\<[10]%
\>[10]{}\mathbin{:}\;\Conid{Stream}_{\text{P}}\;\Varid{m}\;\Varid{m}\;\Conid{A}\;\Varid{→}\;\Conid{Stream}_{\text{W}}\;\Varid{m}\;0\;\Conid{A}{}\<[E]%
\\
\>[3]{}\hsindent{2}{}\<[5]%
\>[5]{}{\anonymous \mkern1mu\constructor{∷}\mkern-1mu\anonymous }\;{}\<[10]%
\>[10]{}\mathbin{:}\;\Conid{A}\;\Varid{→}\;\Conid{Stream}_{\text{W}}\;\Varid{m}\;\Varid{n}\;\Conid{A}\;\Varid{→}\;\Conid{Stream}_{\text{W}}\;\Varid{m}\;(\constructor{suc}\;\Varid{n})\;\Conid{A}{}\<[E]%
\\[\blanklineskip]%
\>[3]{}\Varid{map}_{\text{W}}\;\mathbin{:}\;(\Conid{A}\;\Varid{→}\;\Conid{B})\;\Varid{→}\;\Conid{Stream}_{\text{W}}\;\Varid{m}\;\Varid{n}\;\Conid{A}\;\Varid{→}\;\Conid{Stream}_{\text{W}}\;\Varid{m}\;\Varid{n}\;\Conid{B}{}\<[E]%
\\
\>[3]{}\Varid{map}_{\text{W}}\;\Varid{f}\;[\mskip1.5mu \Varid{xs}\mskip1.5mu]\;{}\<[20]%
\>[20]{}\mathrel{=}\;[\mskip1.5mu \constructor{map}\;\Varid{f}\;\Varid{xs}\mskip1.5mu]{}\<[E]%
\\
\>[3]{}\Varid{map}_{\text{W}}\;\Varid{f}\;(\Varid{x}\;\constructor{∷}\;\Varid{xs})\;{}\<[20]%
\>[20]{}\mathrel{=}\;\Varid{f}\;\Varid{x}\;\constructor{∷}\;\Varid{map}_{\text{W}}\;\Varid{f}\;\Varid{xs}{}\<[E]%
\\[\blanklineskip]%
\>[3]{}\Varid{whnf}\;\mathbin{:}\;\Conid{Stream}_{\text{P}}\;(\constructor{suc}\;\Varid{m})\;\Varid{n}\;\Conid{A}\;\Varid{→}\;\Conid{Stream}_{\text{W}}\;(\constructor{suc}\;\Varid{m})\;\Varid{n}\;\Conid{A}{}\<[E]%
\\
\>[3]{}\Varid{whnf}\;[\mskip1.5mu \Varid{xs}\mskip1.5mu]\;{}\<[21]%
\>[21]{}\mathrel{=}\;[\mskip1.5mu \raisebox{0.8ex}[2ex]{\footnotesize ♭}\;\Varid{xs}\mskip1.5mu]{}\<[E]%
\\
\>[3]{}\Varid{whnf}\;(\Varid{x}\;\constructor{∷}\;\Varid{xs})\;{}\<[21]%
\>[21]{}\mathrel{=}\;\Varid{x}\;\constructor{∷}\;\Varid{whnf}\;\Varid{xs}{}\<[E]%
\\
\>[3]{}\Varid{whnf}\;(\constructor{map}\;\Varid{f}\;\Varid{xs})\;{}\<[21]%
\>[21]{}\mathrel{=}\;\Varid{map}_{\text{W}}\;\Varid{f}\;(\Varid{whnf}\;\Varid{xs}){}\<[E]%
\ColumnHook
\end{hscode}\resethooks
Stream programs where all chunks are non-empty can then be turned into
streams using guarded corecursion:
\begin{hscode}\SaveRestoreHook
\column{B}{@{}>{\hspre}l<{\hspost}@{}}%
\column{3}{@{}>{\hspre}l<{\hspost}@{}}%
\column{5}{@{}>{\hspre}l<{\hspost}@{}}%
\column{20}{@{}>{\hspre}l<{\hspost}@{}}%
\column{24}{@{}>{\hspre}l<{\hspost}@{}}%
\column{E}{@{}>{\hspre}l<{\hspost}@{}}%
\>[3]{}\mathbf{mutual}{}\<[E]%
\\
\>[3]{}\hsindent{2}{}\<[5]%
\>[5]{}\Varid{⟦}\mkern-1mu\anonymous \mkern1mu\Varid{⟧}_{\text{W}}\;\mathbin{:}\;\Conid{Stream}_{\text{W}}\;(\constructor{suc}\;\Varid{m})\;(\constructor{suc}\;\Varid{n})\;\Conid{A}\;\Varid{→}\;\Conid{Stream}\;\Conid{A}{}\<[E]%
\\
\>[3]{}\hsindent{2}{}\<[5]%
\>[5]{}\Varid{⟦}\;\Varid{x}\;\constructor{∷}\;[\mskip1.5mu \Varid{xs}\mskip1.5mu]\;{}\<[20]%
\>[20]{}\Varid{⟧}_{\text{W}}\;{}\<[24]%
\>[24]{}\mathrel{=}\;\Varid{x}\;\constructor{∷}\;\raisebox{0.8ex}[2ex]{\footnotesize ♯}\;\Varid{⟦}\;\Varid{xs}\;\Varid{⟧}_{\text{P}}{}\<[E]%
\\
\>[3]{}\hsindent{2}{}\<[5]%
\>[5]{}\Varid{⟦}\;\Varid{x}\;\constructor{∷}\;(\Varid{y}\;\constructor{∷}\;\Varid{xs})\;\Varid{⟧}_{\text{W}}\;{}\<[24]%
\>[24]{}\mathrel{=}\;\Varid{x}\;\constructor{∷}\;\raisebox{0.8ex}[2ex]{\footnotesize ♯}\;\Varid{⟦}\;\Varid{y}\;\constructor{∷}\;\Varid{xs}\;\Varid{⟧}_{\text{W}}{}\<[E]%
\\[\blanklineskip]%
\>[3]{}\hsindent{2}{}\<[5]%
\>[5]{}\Varid{⟦}\mkern-1mu\anonymous \mkern1mu\Varid{⟧}_{\text{P}}\;\mathbin{:}\;\Conid{Stream}_{\text{P}}\;(\constructor{suc}\;\Varid{m})\;(\constructor{suc}\;\Varid{n})\;\Conid{A}\;\Varid{→}\;\Conid{Stream}\;\Conid{A}{}\<[E]%
\\
\>[3]{}\hsindent{2}{}\<[5]%
\>[5]{}\Varid{⟦}\;\Varid{xs}\;\Varid{⟧}_{\text{P}}\;\mathrel{=}\;\Varid{⟦}\;\Varid{whnf}\;\Varid{xs}\;\Varid{⟧}_{\text{W}}{}\<[E]%
\ColumnHook
\end{hscode}\resethooks
Using this language we cannot define \ensuremath{\Varid{nats}_{\text{2}}}. The following code is
ill-typed:
\begin{hscode}\SaveRestoreHook
\column{B}{@{}>{\hspre}l<{\hspost}@{}}%
\column{3}{@{}>{\hspre}l<{\hspost}@{}}%
\column{E}{@{}>{\hspre}l<{\hspost}@{}}%
\>[3]{}\Varid{nats}_{\text{2}}\;\mathbin{:}\;\Conid{Stream}_{\text{P}}\;\text{2}\;1\;\Conid{ℕ}{}\<[E]%
\\
\>[3]{}\Varid{nats}_{\text{2}}\;\mathrel{=}\;0\;\constructor{∷}\;[\mskip1.5mu \raisebox{0.8ex}[2ex]{\footnotesize ♯}\;\Varid{map}\;\constructor{suc}\;\Varid{nats}_{\text{2}}\mskip1.5mu]{}\<[E]%
\ColumnHook
\end{hscode}\resethooks
On the other hand, the following definitions are accepted:\\
\begin{minipage}{0.5\linewidth}
  \begin{hscode}\SaveRestoreHook
\column{B}{@{}>{\hspre}l<{\hspost}@{}}%
\column{3}{@{}>{\hspre}l<{\hspost}@{}}%
\column{E}{@{}>{\hspre}l<{\hspost}@{}}%
\>[3]{}\Varid{nats}\;\mathbin{:}\;\Conid{Stream}_{\text{P}}\;1\;1\;\Conid{ℕ}{}\<[E]%
\\
\>[3]{}\Varid{nats}\;\mathrel{=}\;0\;\constructor{∷}\;[\mskip1.5mu \raisebox{0.8ex}[2ex]{\footnotesize ♯}\;\Varid{map}\;\constructor{suc}\;\Varid{nats}\mskip1.5mu]{}\<[E]%
\ColumnHook
\end{hscode}\resethooks
\end{minipage}
\begin{minipage}{0.49\linewidth}
  \begin{hscode}\SaveRestoreHook
\column{B}{@{}>{\hspre}l<{\hspost}@{}}%
\column{3}{@{}>{\hspre}l<{\hspost}@{}}%
\column{E}{@{}>{\hspre}l<{\hspost}@{}}%
\>[3]{}\Varid{nats}_{\text{2}}′\;\mathbin{:}\;\Conid{Stream}_{\text{P}}\;\text{2}\;\text{2}\;\Conid{ℕ}{}\<[E]%
\\
\>[3]{}\Varid{nats}_{\text{2}}′\;\mathrel{=}\;0\;\constructor{∷}\;1\;\constructor{∷}\;[\mskip1.5mu \raisebox{0.8ex}[2ex]{\footnotesize ♯}\;\Varid{map}\;\constructor{suc}\;\Varid{nats}_{\text{2}}′\mskip1.5mu]{}\<[E]%
\ColumnHook
\end{hscode}\resethooks
\end{minipage}

The language above uses constant chunk sizes (with the possible
exception of the first chunk).
If more flexibility is needed, then one can index programs by
chunk sizes:\\
\begin{minipage}{0.37\linewidth}
\begin{hscode}\SaveRestoreHook
\column{B}{@{}>{\hspre}l<{\hspost}@{}}%
\column{3}{@{}>{\hspre}l<{\hspost}@{}}%
\column{5}{@{}>{\hspre}l<{\hspost}@{}}%
\column{11}{@{}>{\hspre}l<{\hspost}@{}}%
\column{16}{@{}>{\hspre}l<{\hspost}@{}}%
\column{E}{@{}>{\hspre}l<{\hspost}@{}}%
\>[3]{}\mathbf{data}\;\Conid{Chunks}\;\mathbin{:}\;\Conid{Set}\;\mathbf{where}{}\<[E]%
\\
\>[3]{}\hsindent{2}{}\<[5]%
\>[5]{}\constructor{next}\;{}\<[11]%
\>[11]{}\mathbin{:}\;{}\<[16]%
\>[16]{}\Conid{Chunks}\;\Varid{→}\;\Conid{Chunks}{}\<[E]%
\\
\>[3]{}\hsindent{2}{}\<[5]%
\>[5]{}\constructor{cons}\;{}\<[11]%
\>[11]{}\mathbin{:}\;\Varid{∞}\;{}\<[16]%
\>[16]{}\Conid{Chunks}\;\Varid{→}\;\Conid{Chunks}{}\<[E]%
\ColumnHook
\end{hscode}\resethooks
\end{minipage}
\begin{minipage}{0.62\linewidth}
\begin{hscode}\SaveRestoreHook
\column{B}{@{}>{\hspre}l<{\hspost}@{}}%
\column{3}{@{}>{\hspre}l<{\hspost}@{}}%
\column{5}{@{}>{\hspre}l<{\hspost}@{}}%
\column{10}{@{}>{\hspre}l<{\hspost}@{}}%
\column{E}{@{}>{\hspre}l<{\hspost}@{}}%
\>[3]{}\mathbf{data}\;\Conid{Stream}_{\text{P}}\;\mathbin{:}\;\Conid{Chunks}\;\Varid{→}\;\Conid{Set}\;\Varid{→}\;\Conid{Set}_{\text{1}}\;\mathbf{where}{}\<[E]%
\\
\>[3]{}\hsindent{2}{}\<[5]%
\>[5]{}[\mskip1.5mu \anonymous \mskip1.5mu]\;{}\<[10]%
\>[10]{}\mathbin{:}\;\Varid{∞}\;(\Conid{Stream}_{\text{P}}\;\Varid{cs}\;\Conid{A})\;\Varid{→}\;\Conid{Stream}_{\text{P}}\;(\constructor{next}\;\Varid{cs})\;\Conid{A}{}\<[E]%
\\
\>[3]{}\hsindent{2}{}\<[5]%
\>[5]{}{\anonymous \mkern1mu\constructor{∷}\mkern-1mu\anonymous }\;{}\<[10]%
\>[10]{}\mathbin{:}\;\Conid{A}\;\Varid{→}\;\Conid{Stream}_{\text{P}}\;(\raisebox{0.8ex}[2ex]{\footnotesize ♭}\;\Varid{cs})\;\Conid{A}\;\Varid{→}\;\Conid{Stream}_{\text{P}}\;(\constructor{cons}\;\Varid{cs})\;\Conid{A}{}\<[E]%
\\
\>[3]{}\hsindent{2}{}\<[5]%
\>[5]{}\ldots{}\<[E]%
\ColumnHook
\end{hscode}\resethooks
\end{minipage}
Here \ensuremath{\Conid{Chunks}} represents the chunk sizes used in the production of a
stream: \ensuremath{\constructor{next}} stands for the start of a new chunk, and \ensuremath{\constructor{cons}}
increases the size of the current chunk by one.
Note that \ensuremath{\constructor{next}} is inductive and \ensuremath{\constructor{cons}} coinductive; this ensures
that there are no infinite sequences of empty chunks.

\citet{endrullis-et-al} point out that some approaches to productivity
based on restricted forms of \emph{moduli of production}---which are
closely related to chunk sizes---cannot handle the following
definition of the Thue-Morse sequence:
\begin{hscode}\SaveRestoreHook
\column{B}{@{}>{\hspre}l<{\hspost}@{}}%
\column{3}{@{}>{\hspre}l<{\hspost}@{}}%
\column{E}{@{}>{\hspre}l<{\hspost}@{}}%
\>[3]{}\textit{thue-morse}\;\mathbin{:}\;\Conid{Stream}\;\Conid{Bool}{}\<[E]%
\\
\>[3]{}\textit{thue-morse}\;\mathrel{=}\;\constructor{false}\;\constructor{∷}\;\raisebox{0.8ex}[2ex]{\footnotesize ♯}\;(\Varid{map}\;\Varid{not}\;(\Varid{evens}\;\textit{thue-morse})\;\Varid{⋎}\;\Varid{tail}\;\textit{thue-morse}){}\<[E]%
\ColumnHook
\end{hscode}\resethooks
Here \ensuremath{\Varid{evens}\;\Varid{xs}} consists of every other element of \ensuremath{\Varid{xs}}, starting with
the first, and \ensuremath{{\anonymous \mkern-1mu\Varid{⋎}\mkern-3mu\anonymous }} interleaves two streams: \ensuremath{(\Varid{x}\;\constructor{∷}\;\Varid{xs})\;\Varid{⋎}\;\Varid{ys}\;\mathrel{=}\;\Varid{x}\;\constructor{∷}\;\raisebox{0.8ex}[2ex]{\footnotesize ♯}\;(\Varid{ys}\;\Varid{⋎}\;\raisebox{0.8ex}[2ex]{\footnotesize ♭}\;\Varid{xs})}.
This definition of \ensuremath{\textit{thue-morse}} can be handled using programs indexed
by \ensuremath{\Conid{Chunks}}; see \citet{accompanying-code-par-2010} for details.

\section{Nested Applications}
\label{sec:nested-applications}

Before wrapping up, let us briefly consider nested applications of the
function being defined, as in \ensuremath{\Varid{φ}\;(\Varid{x}\;\constructor{∷}\;\Varid{xs})\;\mathrel{=}\;\Varid{x}\;\constructor{∷}\;\raisebox{0.8ex}[2ex]{\footnotesize ♯}\;\Varid{φ}\;(\Varid{φ}\;\Varid{xs})}.
Definitions with nested applications are common when programs are
written using continuation-passing style.
To handle such applications one can include a constructor for the
function in the type of programs:\\
\begin{minipage}{0.5\linewidth}
\begin{hscode}\SaveRestoreHook
\column{B}{@{}>{\hspre}l<{\hspost}@{}}%
\column{3}{@{}>{\hspre}l<{\hspost}@{}}%
\column{5}{@{}>{\hspre}l<{\hspost}@{}}%
\column{10}{@{}>{\hspre}l<{\hspost}@{}}%
\column{E}{@{}>{\hspre}l<{\hspost}@{}}%
\>[3]{}\mathbf{data}\;\Conid{Stream}_{\text{P}}\;(\Conid{A}\;\mathbin{:}\;\Conid{Set})\;\mathbin{:}\;\Conid{Set}\;\mathbf{where}{}\<[E]%
\\
\>[3]{}\hsindent{2}{}\<[5]%
\>[5]{}{\anonymous \mkern1mu\constructor{∷}\mkern-1mu\anonymous }\;{}\<[10]%
\>[10]{}\mathbin{:}\;\Conid{A}\;\Varid{→}\;\Varid{∞}\;(\Conid{Stream}_{\text{P}}\;\Conid{A})\;\Varid{→}\;\Conid{Stream}_{\text{P}}\;\Conid{A}{}\<[E]%
\\
\>[3]{}\hsindent{2}{}\<[5]%
\>[5]{}\constructor{φ_P}\;{}\<[10]%
\>[10]{}\mathbin{:}\;\Conid{Stream}_{\text{P}}\;\Conid{A}\;\Varid{→}\;\Conid{Stream}_{\text{P}}\;\Conid{A}{}\<[E]%
\\[\blanklineskip]%
\>[3]{}\mathbf{data}\;\Conid{Stream}_{\text{W}}\;(\Conid{A}\;\mathbin{:}\;\Conid{Set})\;\mathbin{:}\;\Conid{Set}\;\mathbf{where}{}\<[E]%
\\
\>[3]{}\hsindent{2}{}\<[5]%
\>[5]{}{\anonymous \mkern1mu\constructor{∷}\mkern-1mu\anonymous }\;\mathbin{:}\;\Conid{A}\;\Varid{→}\;\Conid{Stream}_{\text{P}}\;\Conid{A}\;\Varid{→}\;\Conid{Stream}_{\text{W}}\;\Conid{A}{}\<[E]%
\ColumnHook
\end{hscode}\resethooks
\end{minipage}
\begin{minipage}{0.49\linewidth}
\begin{hscode}\SaveRestoreHook
\column{B}{@{}>{\hspre}l<{\hspost}@{}}%
\column{3}{@{}>{\hspre}l<{\hspost}@{}}%
\column{7}{@{}>{\hspre}l<{\hspost}@{}}%
\column{19}{@{}>{\hspre}l<{\hspost}@{}}%
\column{E}{@{}>{\hspre}l<{\hspost}@{}}%
\>[3]{}\constructor{φ_W}\;{}\<[7]%
\>[7]{}\mathbin{:}\;\Conid{Stream}_{\text{W}}\;\Conid{A}\;\Varid{→}\;\Conid{Stream}_{\text{W}}\;\Conid{A}{}\<[E]%
\\
\>[3]{}\constructor{φ_W}\;(\Varid{x}\;\constructor{∷}\;\Varid{xs})\;\mathrel{=}\;\Varid{x}\;\constructor{∷}\;\constructor{φ_P}\;(\constructor{φ_P}\;\Varid{xs}){}\<[E]%
\\[\blanklineskip]%
\>[3]{}\Varid{whnf}\;\mathbin{:}\;\Conid{Stream}_{\text{P}}\;\Conid{A}\;\Varid{→}\;\Conid{Stream}_{\text{W}}\;\Conid{A}{}\<[E]%
\\
\>[3]{}\Varid{whnf}\;(\Varid{x}\;\constructor{∷}\;\Varid{xs})\;{}\<[19]%
\>[19]{}\mathrel{=}\;\Varid{x}\;\constructor{∷}\;\raisebox{0.8ex}[2ex]{\footnotesize ♭}\;\Varid{xs}{}\<[E]%
\\
\>[3]{}\Varid{whnf}\;(\constructor{φ_P}\;\Varid{xs})\;{}\<[19]%
\>[19]{}\mathrel{=}\;\constructor{φ_W}\;(\Varid{whnf}\;\Varid{xs}){}\<[E]%
\ColumnHook
\end{hscode}\resethooks
\end{minipage}
(The definition of \ensuremath{\Varid{⟦}\mkern-1mu\anonymous \mkern1mu\Varid{⟧}_{\text{P}}} is omitted above.)
By turning streams into programs one can then define \ensuremath{\Varid{φ}}:\\
\begin{minipage}{0.5\linewidth}
  \begin{hscode}\SaveRestoreHook
\column{B}{@{}>{\hspre}l<{\hspost}@{}}%
\column{5}{@{}>{\hspre}l<{\hspost}@{}}%
\column{E}{@{}>{\hspre}l<{\hspost}@{}}%
\>[5]{}{\Varid{⌈}\mkern-2mu\anonymous \mkern-0.5mu\Varid{⌉}}\;\mathbin{:}\;\Conid{Stream}\;\Conid{A}\;\Varid{→}\;\Conid{Stream}_{\text{P}}\;\Conid{A}{}\<[E]%
\\
\>[5]{}\Varid{⌈}\;\Varid{x}\;\constructor{∷}\;\Varid{xs}\;\Varid{⌉}\;\mathrel{=}\;\Varid{x}\;\constructor{∷}\;\raisebox{0.8ex}[2ex]{\footnotesize ♯}\;\Varid{⌈}\;\raisebox{0.8ex}[2ex]{\footnotesize ♭}\;\Varid{xs}\;\Varid{⌉}{}\<[E]%
\ColumnHook
\end{hscode}\resethooks
\end{minipage}
\begin{minipage}{0.49\linewidth}
  \begin{hscode}\SaveRestoreHook
\column{B}{@{}>{\hspre}l<{\hspost}@{}}%
\column{5}{@{}>{\hspre}l<{\hspost}@{}}%
\column{E}{@{}>{\hspre}l<{\hspost}@{}}%
\>[5]{}\Varid{φ}\;\mathbin{:}\;\Conid{Stream}\;\Conid{A}\;\Varid{→}\;\Conid{Stream}\;\Conid{A}{}\<[E]%
\\
\>[5]{}\Varid{φ}\;\Varid{xs}\;\mathrel{=}\;\Varid{⟦}\;\constructor{φ_P}\;\Varid{⌈}\;\Varid{xs}\;\Varid{⌉}\;\Varid{⟧}_{\text{P}}{}\<[E]%
\ColumnHook
\end{hscode}\resethooks
\end{minipage}
In order to prove that \ensuremath{\Varid{φ}} satisfies its intended defining equation it
can be helpful to use an equality proof language, as in
Section~\ref{sec:proofs}, and to include a constructor for the
congruence of \ensuremath{\constructor{φ_P}} in this language:
\begin{hscode}\SaveRestoreHook
\column{B}{@{}>{\hspre}l<{\hspost}@{}}%
\column{3}{@{}>{\hspre}l<{\hspost}@{}}%
\column{5}{@{}>{\hspre}l<{\hspost}@{}}%
\column{16}{@{}>{\hspre}l<{\hspost}@{}}%
\column{E}{@{}>{\hspre}l<{\hspost}@{}}%
\>[3]{}\mathbf{data}\;{\anonymous \mkern-4mu≈_{\text{P}}\mkern-6mu\anonymous }\;\mathbin{:}\;\Conid{Stream}\;\Conid{A}\;\Varid{→}\;\Conid{Stream}\;\Conid{A}\;\Varid{→}\;\Conid{Set}\;\mathbf{where}{}\<[E]%
\\
\>[3]{}\hsindent{2}{}\<[5]%
\>[5]{}\ldots{}\<[E]%
\\
\>[3]{}\hsindent{2}{}\<[5]%
\>[5]{}\constructor{φ_P}\textit{-}\constructor{cong}\;\mathbin{:}\;{}\<[16]%
\>[16]{}(\Varid{xs}\;\Varid{ys}\;\mathbin{:}\;\Conid{Stream}_{\text{P}}\;\Conid{A})\;\Varid{→}\;\Varid{⟦}\;\Varid{xs}\;\Varid{⟧}_{\text{P}}\;≈_{\text{P}}\;\Varid{⟦}\;\Varid{ys}\;\Varid{⟧}_{\text{P}}\;\Varid{→}\;\Varid{⟦}\;\constructor{φ_P}\;\Varid{xs}\;\Varid{⟧}_{\text{P}}\;≈_{\text{P}}\;\Varid{⟦}\;\constructor{φ_P}\;\Varid{ys}\;\Varid{⟧}_{\text{P}}{}\<[E]%
\ColumnHook
\end{hscode}\resethooks
For further details, see \citet{accompanying-code-par-2010}, who also
establishes that \ensuremath{\Varid{φ}}'s defining equation has a unique solution.

\section{Related Work}
\label{sec:related-work}

This section is mainly concerned with discussing methods for
establishing productivity \emph{in systems based on guarded
  corecursion}. Other related work is discussed towards the end.

\citet{rutten} proves that certain operations on streams are
well-defined by using a technique which is very similar to the one
described in this paper. He defines a language \ensuremath{\Conid{E}} of real number
stream expressions inductively (this language is similar to \ensuremath{\Conid{Stream}_{\text{P}}\;\Conid{ℝ}}), and defines a stream coalgebra \ensuremath{\Varid{c}\;\mathbin{:}\;\Conid{E}\;\Varid{→}\;\Conid{ℝ}\;\Varid{×}\;\Conid{E}} by recursion over
the structure of \ensuremath{\Conid{E}} (this corresponds to \ensuremath{\Varid{whnf}}). The type of streams
is a final coalgebra, so from \ensuremath{\Varid{c}} one obtains a function of type \ensuremath{\Conid{E}\;\Varid{→}\;\Conid{Stream}\;\Conid{ℝ}} (corresponding to \ensuremath{\Varid{⟦}\mkern-1mu\anonymous \mkern1mu\Varid{⟧}_{\text{P}}}), which can be used to turn stream
expressions into actual streams. Rutten then uses coinduction
(expressed using bisimulations) to prove that the defined operations
satisfy their intended defining equations, and that these equations
have unique solutions.

There are some differences between Rutten's proof and the technique
described here, other than the different settings (finality vs.\@
guarded corecursion, bisimulations vs.\@ guarded coinduction).
One is that Rutten defines the variant of \ensuremath{\Varid{fib}} from
Section~\ref{sec:destructors} via two mutually recursive streams (\ensuremath{\Varid{fib}\;\mathrel{=}\;0\;\constructor{∷}\;\raisebox{0.8ex}[2ex]{\footnotesize ♯}\;\Varid{fib'}} and \ensuremath{\Varid{fib'}\;\mathrel{=}\;1\;\constructor{∷}\;\raisebox{0.8ex}[2ex]{\footnotesize ♯}\;\Varid{zip\mkern-2mu{}With}\;{\anonymous \mkern-0.5mu\Varid{+}\mkern-2mu\anonymous \mkern3mu}\;\Varid{fib}\;\Varid{fib'}}); he does not
discuss anything resembling the counting approaches of
Sections~\ref{sec:destructors} and~\ref{sec:other-chunk-sizes}.
Another difference is that Rutten's language \ensuremath{\Conid{E}} is inductive, whereas
\ensuremath{\Conid{Stream}_{\text{P}}} uses mixed induction and coinduction.
A simple consequence of this difference is that when Rutten defines
\ensuremath{\Varid{fib}} he includes it as a term in \ensuremath{\Conid{E}}; with the method described here
one can get much further using a fixed language.
\citet{danielsson-altenkirch-subtyping}
also take advantage of this difference when proving that one subtyping
relation is sound with respect to another.
In this proof the program and WHNF types are defined mutually, using
mixed induction and coinduction, and the \ensuremath{\Varid{whnf}} function constructs
its result using a combination of structural recursion and guarded
corecursion.
For completeness a short variant of this development is included in
Appendix~\ref{sec:inductive-stream-equality}.

Rutten's proof is closely related to a technique due to
\citet{bartels}.
Bartels formulates the technique in a general categorical setting,
and restricts the form of \ensuremath{\Varid{whnf}}, and in return proofs showing that
the definitions uniquely satisfy certain defining equations come for
free.
Furthermore Bartels manages to define \ensuremath{\Varid{fib}} without including it as a
term in the language.

\citet{niqui,niqui-tr} implements one of Bartels' corecursion schemes,
λ-coiteration, in Coq. He states that this scheme cannot handle
van~de~Snepscheut's corecursive definition of the Hamming numbers
\citep{EWD792}, which can easily be handled using the method described
in this paper.

\citet{matthews} and \citet{gianantonio-miculan} describe general
frameworks for defining values using a mixture of recursion and
corecursion, based on functions which satisfy notions of
contractivity. The methods seem to be quite general, and have been
implemented (in Isabelle and Coq, respectively; note that guarded
corecursion is not a primitive feature of Isabelle).

The implementations mentioned above
\citep{matthews,gianantonio-miculan,niqui,niqui-tr} provide you with
unique solutions to equations, whereas when using the method described
in this paper you need to prove correctness and uniqueness manually if
you are interested in these properties.
On the other hand, as pointed out in Section~\ref{sec:proofs}, there
is rarely any need to pay this price when defining a proof.
I suspect that circumstances determine which method is cheapest to
use.

\citet{bertot} implements a filter function for streams in Coq. An
unrestricted filter function is not productive, so Bertot restricts
the function's inputs using predicates of the form ``always
(eventually $P$)''. The always part is defined coinductively, and the
eventually part inductively.
As mentioned in the introduction this work is orthogonal to the work
presented here.

Conor McBride (personal communication) has developed a technique for
establishing productivity, based on the work of
\citet{hancock-setzer}. The idea is to represent the right-hand sides
of function definitions using a type \ensuremath{\Conid{RHS}\;\Varid{g}}, where \ensuremath{\Varid{g}} indicates
whether the context is guarding or not, and to only allow corecursive
calls in a guarding context.

\citet{capretta} defines the partiality monad, which can be used to
represent partial (potentially non-terminating) computations, roughly
as follows:
\begin{hscode}\SaveRestoreHook
\column{B}{@{}>{\hspre}l<{\hspost}@{}}%
\column{3}{@{}>{\hspre}l<{\hspost}@{}}%
\column{5}{@{}>{\hspre}l<{\hspost}@{}}%
\column{12}{@{}>{\hspre}l<{\hspost}@{}}%
\column{23}{@{}>{\hspre}l<{\hspost}@{}}%
\column{E}{@{}>{\hspre}l<{\hspost}@{}}%
\>[3]{}\mathbf{data}\;\anonymous {}^{\nu}\;(\Conid{A}\;\mathbin{:}\;\Conid{Set})\;\mathbin{:}\;\Conid{Set}\;\mathbf{where}{}\<[E]%
\\
\>[3]{}\hsindent{2}{}\<[5]%
\>[5]{}\constructor{return}\;{}\<[12]%
\>[12]{}\mathbin{:}\;\Conid{A}\;{}\<[23]%
\>[23]{}\Varid{→}\;\Conid{A}\;{}^{\nu}{}\<[E]%
\\
\>[3]{}\hsindent{2}{}\<[5]%
\>[5]{}\constructor{step}\;{}\<[12]%
\>[12]{}\mathbin{:}\;\Varid{∞}\;(\Conid{A}\;{}^{\nu})\;{}\<[23]%
\>[23]{}\Varid{→}\;\Conid{A}\;{}^{\nu}{}\<[E]%
\ColumnHook
\end{hscode}\resethooks
The constructor \ensuremath{\constructor{return}} returns a result, and \ensuremath{\constructor{step}} postpones a
computation.
It is easy to define bind for this monad: \ensuremath{{\anonymous \mkern-4mu\bind \mkern-5.5mu\anonymous }\;\mathbin{:}\;\Conid{A}\;{}^{\nu}\;\Varid{→}\;(\Conid{A}\;\Varid{→}\;\Conid{B}\;{}^{\nu})\;\Varid{→}\;\Conid{B}\;{}^{\nu}}.
Unfortunately it can be inconvenient to use this definition of bind in
systems based on guarded corecursion, because \ensuremath{{\anonymous \mkern-4mu\bind \mkern-5.5mu\anonymous }} is not a
constructor.
\citet{megacz} suggests (more or less) the following alternative
definition:
\begin{hscode}\SaveRestoreHook
\column{B}{@{}>{\hspre}l<{\hspost}@{}}%
\column{3}{@{}>{\hspre}l<{\hspost}@{}}%
\column{5}{@{}>{\hspre}l<{\hspost}@{}}%
\column{12}{@{}>{\hspre}l<{\hspost}@{}}%
\column{E}{@{}>{\hspre}l<{\hspost}@{}}%
\>[3]{}\mathbf{data}\;\anonymous {}^{\nu}\;(\Conid{A}\;\mathbin{:}\;\Conid{Set})\;\mathbin{:}\;\Conid{Set}_{\text{1}}\;\mathbf{where}{}\<[E]%
\\
\>[3]{}\hsindent{2}{}\<[5]%
\>[5]{}\constructor{return}\;{}\<[12]%
\>[12]{}\mathbin{:}\;\Conid{A}\;\Varid{→}\;\Conid{A}\;{}^{\nu}{}\<[E]%
\\
\>[3]{}\hsindent{2}{}\<[5]%
\>[5]{}{\anonymous \mkern-4mu\bind \mkern-5.5mu\anonymous }\;{}\<[12]%
\>[12]{}\mathbin{:}\;\Varid{∞}\;(\Conid{B}\;{}^{\nu})\;\Varid{→}\;(\Conid{B}\;\Varid{→}\;\Varid{∞}\;(\Conid{A}\;{}^{\nu}))\;\Varid{→}\;\Conid{A}\;{}^{\nu}{}\<[E]%
\ColumnHook
\end{hscode}\resethooks
One can note that this is very close to the first step of the
technique presented in this paper. Megacz does not translate from the
second to the first type, though.

\citet{bertot-komendantskaya-types} describe a method for replacing
corecursion with recursion. They map values of type \ensuremath{\Conid{Stream}\;\Conid{A}} to and
from the isomorphic type \ensuremath{\Conid{ℕ}\;\Varid{→}\;\Conid{A}}, and values of this type can be
defined recursively.
The authors state that the method is still very limited and that, as
presented, it cannot handle van~de~Snepscheut's definition of the
Hamming numbers.

\citet{mcbride} defines an applicative functor which captures the
notion of ``be[ing] ready a wee bit later''. Using this structure he
defines various corecursive programs, including the circular
breadth-first labelling function which is defined in
Section~\ref{sec:multiple-types}.
The technique is presented using the partial language Haskell, but
Robert Atkey (personal communication) has later implemented it in
Agda.
The technique has not been developed very far yet: as far as I am
aware no one has tried to prove any properties about functions defined
using it.

Instead of working around the limitations of guarded corecursion one
can include language features which make it easier to explain why
programs are productive. One such feature is \emph{sized types}
\citep{hughes-pareto-sabry,barthe-et-al,abel}, and the λ-calculi of
\citet{buchholz} provide other examples.
Another approach is to use cleverer algorithms for establishing
productivity.
\citet{endrullis-et-al,endrullis-grabmayer-hendriks} present
algorithms which handle the definition of \ensuremath{\textit{thue-morse}} from
Section~\ref{sec:other-chunk-sizes} automatically (except that, as
presented, they only support first-order term-rewriting systems).
The algorithms are tailored to streams; it seems to be hard to adapt
them to, say, coinductive trees.
Another algorithm is presented by \citet{telford-turner}.
This algorithm does not handle \ensuremath{\textit{thue-morse}} \citep{endrullis-et-al},
but has the advantage of working for a large class of coinductive data
types.

\citet{morris-altenkirch-mcbride} use the technique of replacing
functions with constructors to show \emph{termination} rather than
productivity (see \citet{morris-altenkirch-ghani} for an explanation
of the technique). They replace a partially applied recursive call
(which is not necessarily structural, because it could later be
applied to anything), nested inside another recursive call, with a
constructor application. If this constructor application is later
encountered it is handled using structural recursion.

The technique presented here also shares some traits with
\citeauthor{reynolds}' \emph{defunctionalisation}
\citeyearpar{reynolds}. Defunctionalisation is used to translate
programs written in higher-order languages to first-order languages,
and it basically amounts to representing function spaces using
application-specific data types, and implementing interpreters for
these data types.

\section{Conclusions}
\label{sec:conclusions}

I hope to have shown, through a number of examples, that the
language-based approach to establishing productivity is useful.
I am currently turning to it whenever I have a problem with
guardedness; see \citet{danielsson-altenkirch-subtyping} and
\citet{danielsson-parser-combinators} for some examples not included
in this paper.

However, there are some problems with the method. As discussed in
Section~\ref{sec:proofs} it is not very useful if efficiency is a
concern. Furthermore it can be disruptive: if one decides to use the
method after already having developed a large number of functions in
some project, and many of these functions have to be reified as
constructors in a program data type, then a lot of work may be
necessary.
In fact, this problem---in one shape or another---is likely to apply
to \emph{all} approaches to making definitions guarded.
In the long term I believe that it would be useful to adopt a more
modular approach to productivity than guardedness.

\paragraph{Acknowledgements.} I would like to thank Andreas Abel,
Thorsten Altenkirch, Conor McBride, Nicolas Oury and Anton Setzer for
many discussions about coinduction, and Graham Hutton as well as
several anonymous reviewers for useful feedback. I would also like to
thank EPSRC for financial support (grant code: EP/E04350X/1).

\appendix

\section{An Inductive Approximation of Stream Equality}
\label{sec:inductive-stream-equality}

\citet{danielsson-altenkirch-subtyping} prove that one subtyping
relation is sound with respect to another using the technique
described in this paper. This appendix outlines the proof, but in a
simplified (and slightly different) setting: equality between streams.

Recall the definitions of \ensuremath{\Conid{Stream}} and stream equality, \ensuremath{{\anonymous \mkern-4mu≈\mkern-6mu\anonymous }}, from
Section~\ref{sec:coinduction}. One can define a sound
ap-\linebreak{}proximation of stream equality inductively as follows
(using an idea due to \citet{brandt-henglein-fi}):
\begin{hscode}\SaveRestoreHook
\column{B}{@{}>{\hspre}l<{\hspost}@{}}%
\column{3}{@{}>{\hspre}l<{\hspost}@{}}%
\column{11}{@{}>{\hspre}l<{\hspost}@{}}%
\column{E}{@{}>{\hspre}l<{\hspost}@{}}%
\>[B]{}\mathbf{data}\;{\anonymous \mkern-3mu\mathrel{⊢}\mkern-6mu{\anonymous \mkern-4mu≈\mkern-6mu\anonymous }}\;(\Conid{H}\;\mathbin{:}\;\Conid{List}\;(\Conid{Stream}\;\Conid{A}\;\Varid{×}\;\Conid{Stream}\;\Conid{A}))\;\mathbin{:}\;\Conid{Stream}\;\Conid{A}\;\Varid{→}\;\Conid{Stream}\;\Conid{A}\;\Varid{→}\;\Conid{Set}\;\mathbf{where}{}\<[E]%
\\
\>[B]{}\hsindent{3}{}\<[3]%
\>[3]{}{\anonymous \mkern1mu\constructor{∷}\mkern-1mu\anonymous }\;{}\<[11]%
\>[11]{}\mathbin{:}\;(\Varid{x}\;\mathbin{:}\;\Conid{A})\;\ \to\ \;(\Varid{x}\;\constructor{∷}\;\Varid{xs},\Varid{x}\;\constructor{∷}\;\Varid{ys})\;\constructor{∷}\;\Conid{H}\;\mathrel{⊢}\;\raisebox{0.8ex}[2ex]{\footnotesize ♭}\;\Varid{xs}\;≈\;\raisebox{0.8ex}[2ex]{\footnotesize ♭}\;\Varid{ys}\;\ \to\ \;\Conid{H}\;\mathrel{⊢}\;\Varid{x}\;\constructor{∷}\;\Varid{xs}\;≈\;\Varid{x}\;\constructor{∷}\;\Varid{ys}{}\<[E]%
\\
\>[B]{}\hsindent{3}{}\<[3]%
\>[3]{}\constructor{hyp}\;{}\<[11]%
\>[11]{}\mathbin{:}\;(\Varid{xs},\Varid{ys})\;\mathrel{∈}\;\Conid{H}\;\ \to\ \;\Conid{H}\;\mathrel{⊢}\;\Varid{xs}\;≈\;\Varid{ys}{}\<[E]%
\\
\>[B]{}\hsindent{3}{}\<[3]%
\>[3]{}\constructor{trans}\;{}\<[11]%
\>[11]{}\mathbin{:}\;\Conid{H}\;\mathrel{⊢}\;\Varid{xs}\;≈\;\Varid{ys}\;\ \to\ \;\Conid{H}\;\mathrel{⊢}\;\Varid{ys}\;≈\;\Varid{zs}\;\ \to\ \;\Conid{H}\;\mathrel{⊢}\;\Varid{xs}\;≈\;\Varid{zs}{}\<[E]%
\ColumnHook
\end{hscode}\resethooks
The intention is that, if one can prove \ensuremath{\Conid{H}\;\mathrel{⊢}\;\Varid{xs}\;≈\;\Varid{ys}}, and all the
assumptions in the list \ensuremath{\Conid{H}} are valid, then \ensuremath{\Varid{xs}} and \ensuremath{\Varid{ys}} should be
equal.
The first constructor of \ensuremath{{\anonymous \mkern-3mu\mathrel{⊢}\mkern-6mu{\anonymous \mkern-4mu≈\mkern-6mu\anonymous }}} states that, in order to prove that
\ensuremath{\Varid{x}\;\constructor{∷}\;\Varid{xs}} and \ensuremath{\Varid{x}\;\constructor{∷}\;\Varid{ys}} are equal, it suffices to show that \ensuremath{\raisebox{0.8ex}[2ex]{\footnotesize ♭}\;\Varid{xs}} and
\ensuremath{\raisebox{0.8ex}[2ex]{\footnotesize ♭}\;\Varid{ys}} are equal, given the extra assumption that \ensuremath{\Varid{x}\;\constructor{∷}\;\Varid{xs}} and \ensuremath{\Varid{x}\;\constructor{∷}\;\Varid{ys}} are equal.
The second constructor makes it possible to use the hypotheses in the
list \ensuremath{\Conid{H}} (\ensuremath{{\anonymous \mkern-5mu\mathrel{∈}\mkern-6mu\anonymous }} encodes list membership), and the third constructor
encodes transitivity.
As an example, we can prove that the list \ensuremath{\Varid{repeat}\;\Varid{x}\;≈\;\Varid{x}\;\constructor{∷}\;\raisebox{0.8ex}[2ex]{\footnotesize ♯}\;\Varid{repeat}\;\Varid{x}}
is equal to itself as follows:
\begin{hscode}\SaveRestoreHook
\column{B}{@{}>{\hspre}l<{\hspost}@{}}%
\column{3}{@{}>{\hspre}l<{\hspost}@{}}%
\column{E}{@{}>{\hspre}l<{\hspost}@{}}%
\>[3]{}\textit{repeat-refl}\;\mathbin{:}\;(\Varid{x}\;\mathbin{:}\;\Conid{A})\;\ \to\ \;[\mskip1.5mu \mskip1.5mu]\;\mathrel{⊢}\;\Varid{repeat}\;\Varid{x}\;≈\;\Varid{repeat}\;\Varid{x}{}\<[E]%
\\
\>[3]{}\textit{repeat-refl}\;\Varid{x}\;\mathrel{=}\;\Varid{x}\;\constructor{∷}\;\constructor{hyp}\;\constructor{here}{}\<[E]%
\ColumnHook
\end{hscode}\resethooks
(The constructor \ensuremath{\constructor{here}} proves that the head of a list is a member of
the list. In this case it is used at the type \ensuremath{(\Varid{repeat}\;\Varid{x},\Varid{repeat}\;\Varid{x})\;\mathrel{∈}\;(\Varid{repeat}\;\Varid{x},\Varid{repeat}\;\Varid{x})\;\constructor{∷}\;[\mskip1.5mu \mskip1.5mu]}.)

Soundness of \ensuremath{{\anonymous \mkern-3mu\mathrel{⊢}\mkern-6mu{\anonymous \mkern-4mu≈\mkern-6mu\anonymous }}} will now be established.
The goal is to prove that \ensuremath{\Conid{H}\;\mathrel{⊢}\;\Varid{xs}\;≈\;\Varid{ys}} implies \ensuremath{\Varid{xs}\;≈\;\Varid{ys}}, given that
\ensuremath{\Conid{All}\;(\Conid{Valid}\;{\anonymous \mkern-4mu≈\mkern-6mu\anonymous })\;\Conid{H}}, where \ensuremath{\Conid{All}\;\Conid{P}\;\Varid{xs}} means that \ensuremath{\Conid{P}} holds for all
elements in the list \ensuremath{\Varid{xs}}, and \ensuremath{\Conid{Valid}} is \ensuremath{\Varid{uncurry}} for stream
predicates:
\begin{hscode}\SaveRestoreHook
\column{B}{@{}>{\hspre}l<{\hspost}@{}}%
\column{3}{@{}>{\hspre}l<{\hspost}@{}}%
\column{5}{@{}>{\hspre}l<{\hspost}@{}}%
\column{10}{@{}>{\hspre}l<{\hspost}@{}}%
\column{E}{@{}>{\hspre}l<{\hspost}@{}}%
\>[3]{}\mathbf{data}\;\Conid{All}\;(\Conid{P}\;\mathbin{:}\;\Conid{A}\;\Varid{→}\;\Conid{Set})\;\mathbin{:}\;\Conid{List}\;\Conid{A}\;\Varid{→}\;\Conid{Set}\;\mathbf{where}{}\<[E]%
\\
\>[3]{}\hsindent{2}{}\<[5]%
\>[5]{}[\mskip1.5mu \mskip1.5mu]\;{}\<[10]%
\>[10]{}\mathbin{:}\;\Conid{All}\;\Conid{P}\;[\mskip1.5mu \mskip1.5mu]{}\<[E]%
\\
\>[3]{}\hsindent{2}{}\<[5]%
\>[5]{}{\anonymous \mkern1mu\constructor{∷}\mkern-1mu\anonymous }\;{}\<[10]%
\>[10]{}\mathbin{:}\;\Conid{P}\;\Varid{x}\;\Varid{→}\;\Conid{All}\;\Conid{P}\;\Varid{xs}\;\Varid{→}\;\Conid{All}\;\Conid{P}\;(\Varid{x}\;\constructor{∷}\;\Varid{xs}){}\<[E]%
\\[\blanklineskip]%
\>[3]{}\Conid{Valid}\;\mathbin{:}\;(\Conid{Stream}\;\Conid{A}\;\Varid{→}\;\Conid{Stream}\;\Conid{A}\;\Varid{→}\;\Conid{Set})\;\ \to\ \;\Conid{Stream}\;\Conid{A}\;\Varid{×}\;\Conid{Stream}\;\Conid{A}\;\Varid{→}\;\Conid{Set}{}\<[E]%
\\
\>[3]{}\Conid{Valid}\;{\anonymous \mkern2mu\Conid{R}\mkern-0.7mu\anonymous }\;(\Varid{xs},\Varid{ys})\;\mathrel{=}\;\Varid{xs}\;\Conid{R}\;\Varid{ys}{}\<[E]%
\ColumnHook
\end{hscode}\resethooks
We begin by defining the program and WHNF types mutually as follows:
\begin{hscode}\SaveRestoreHook
\column{B}{@{}>{\hspre}l<{\hspost}@{}}%
\column{3}{@{}>{\hspre}l<{\hspost}@{}}%
\column{5}{@{}>{\hspre}l<{\hspost}@{}}%
\column{13}{@{}>{\hspre}l<{\hspost}@{}}%
\column{E}{@{}>{\hspre}l<{\hspost}@{}}%
\>[B]{}\mathbf{mutual}{}\<[E]%
\\
\>[B]{}\hsindent{3}{}\<[3]%
\>[3]{}\mathbf{data}\;{\anonymous \mkern-4mu≈_{\text{P}}\mkern-6mu\anonymous }\;\mathbin{:}\;\Conid{Stream}\;\Conid{A}\;\Varid{→}\;\Conid{Stream}\;\Conid{A}\;\Varid{→}\;\Conid{Set}\;\mathbf{where}{}\<[E]%
\\
\>[3]{}\hsindent{2}{}\<[5]%
\>[5]{}\constructor{sound}\;{}\<[13]%
\>[13]{}\mathbin{:}\;\Conid{All}\;(\Conid{Valid}\;{\anonymous \mkern-4mu≈_{\text{W}}\mkern-6mu\anonymous })\;\Conid{H}\;\ \to\ \;\Conid{H}\;\mathrel{⊢}\;\Varid{xs}\;≈\;\Varid{ys}\;\ \to\ \;\Varid{xs}\;≈_{\text{P}}\;\Varid{ys}{}\<[E]%
\\
\>[3]{}\hsindent{2}{}\<[5]%
\>[5]{}\constructor{trans}\;{}\<[13]%
\>[13]{}\mathbin{:}\;\Varid{xs}\;≈_{\text{P}}\;\Varid{ys}\;\ \to\ \;\Varid{ys}\;≈_{\text{P}}\;\Varid{zs}\;\ \to\ \;\Varid{xs}\;≈_{\text{P}}\;\Varid{zs}{}\<[E]%
\\[\blanklineskip]%
\>[B]{}\hsindent{3}{}\<[3]%
\>[3]{}\mathbf{data}\;{\anonymous \mkern-4mu≈_{\text{W}}\mkern-6mu\anonymous }\;\mathbin{:}\;\Conid{Stream}\;\Conid{A}\;\Varid{→}\;\Conid{Stream}\;\Conid{A}\;\Varid{→}\;\Conid{Set}\;\mathbf{where}{}\<[E]%
\\
\>[3]{}\hsindent{2}{}\<[5]%
\>[5]{}{\anonymous \mkern1mu\constructor{∷}\mkern-1mu\anonymous }\;\mathbin{:}\;(\Varid{x}\;\mathbin{:}\;\Conid{A})\;\ \to\ \;\Varid{∞}\;(\raisebox{0.8ex}[2ex]{\footnotesize ♭}\;\Varid{xs}\;≈_{\text{P}}\;\raisebox{0.8ex}[2ex]{\footnotesize ♭}\;\Varid{ys})\;\ \to\ \;\Varid{x}\;\constructor{∷}\;\Varid{xs}\;≈_{\text{W}}\;\Varid{x}\;\constructor{∷}\;\Varid{ys}{}\<[E]%
\ColumnHook
\end{hscode}\resethooks
Note that the first argument of the program \ensuremath{\constructor{sound}} refers to WHNFs.
The function \ensuremath{\Varid{trans}_{\text{W}}}, whose type is \ensuremath{\Varid{xs}\;≈_{\text{W}}\;\Varid{ys}\;\Varid{→}\;\Varid{ys}\;≈_{\text{W}}\;\Varid{zs}\;\Varid{→}\;\Varid{xs}\;≈_{\text{W}}\;\Varid{zs}},
can be defined using simple case analysis.
The function \ensuremath{\Varid{sound}_{\text{W}}} is defined as follows, using structural
recursion:
\begin{hscode}\SaveRestoreHook
\column{B}{@{}>{\hspre}l<{\hspost}@{}}%
\column{3}{@{}>{\hspre}l<{\hspost}@{}}%
\column{36}{@{}>{\hspre}l<{\hspost}@{}}%
\column{E}{@{}>{\hspre}l<{\hspost}@{}}%
\>[B]{}\Varid{sound}_{\text{W}}\;\mathbin{:}\;\Conid{All}\;(\Conid{Valid}\;{\anonymous \mkern-4mu≈_{\text{W}}\mkern-6mu\anonymous })\;\Conid{H}\;\ \to\ \;\Conid{H}\;\mathrel{⊢}\;\Varid{xs}\;≈\;\Varid{ys}\;\ \to\ \;\Varid{xs}\;≈_{\text{W}}\;\Varid{ys}{}\<[E]%
\\
\>[B]{}\Varid{sound}_{\text{W}}\;\Varid{valid}\;(\constructor{hyp}\;\Varid{h})\;{}\<[36]%
\>[36]{}\mathrel{=}\;\Varid{lookup}\;\Varid{valid}\;\Varid{h}{}\<[E]%
\\
\>[B]{}\Varid{sound}_{\text{W}}\;\Varid{valid}\;(\constructor{trans}\;\Varid{xs}\mkern-5mu≈\mkern-6mu\Varid{ys}\;\Varid{ys}\mkern-5mu≈\mkern-6mu\Varid{zs})\;{}\<[36]%
\>[36]{}\mathrel{=}\;\Varid{trans}_{\text{W}}\;(\Varid{sound}_{\text{W}}\;\Varid{valid}\;\Varid{xs}\mkern-5mu≈\mkern-6mu\Varid{ys})\;(\Varid{sound}_{\text{W}}\;\Varid{valid}\;\Varid{ys}\mkern-5mu≈\mkern-6mu\Varid{zs}){}\<[E]%
\\
\>[B]{}\Varid{sound}_{\text{W}}\;\Varid{valid}\;(\Varid{x}\;\constructor{∷}\;\Varid{xs}\mkern-5mu≈\mkern-6mu\Varid{ys})\;{}\<[36]%
\>[36]{}\mathrel{=}\;\Varid{proof}{}\<[E]%
\\
\>[B]{}\hsindent{3}{}\<[3]%
\>[3]{}\mathbf{where}\;\Varid{proof}\;\mathrel{=}\;\Varid{x}\;\constructor{∷}\;\raisebox{0.8ex}[2ex]{\footnotesize ♯}\;\constructor{sound}\;(\Varid{proof}\;\constructor{∷}\;\Varid{valid})\;\Varid{xs}\mkern-5mu≈\mkern-6mu\Varid{ys}{}\<[E]%
\ColumnHook
\end{hscode}\resethooks
In the first clause \ensuremath{\Varid{lookup}\;\mathbin{:}\;\Conid{All}\;\Conid{P}\;\Varid{xs}\;\Varid{→}\;\Varid{x}\;\mathrel{∈}\;\Varid{xs}\;\Varid{→}\;\Conid{P}\;\Varid{x}} is used to
fetch a proof from the ``list'' of valid assumptions.
In the third clause a \emph{circular} proof is constructed using
guarded corecursion; note that the list of valid assumptions is
extended with the proof currently being defined.
Given \ensuremath{\Varid{trans}_{\text{W}}} and \ensuremath{\Varid{sound}_{\text{W}}} it is easy to define \ensuremath{\Varid{whnf}} using
structural recursion:
\begin{hscode}\SaveRestoreHook
\column{B}{@{}>{\hspre}l<{\hspost}@{}}%
\column{28}{@{}>{\hspre}l<{\hspost}@{}}%
\column{E}{@{}>{\hspre}l<{\hspost}@{}}%
\>[B]{}\Varid{whnf}\;\mathbin{:}\;\Varid{xs}\;≈_{\text{P}}\;\Varid{ys}\;\Varid{→}\;\Varid{xs}\;≈_{\text{W}}\;\Varid{ys}{}\<[E]%
\\
\>[B]{}\Varid{whnf}\;(\constructor{sound}\;\Varid{valid}\;\Varid{xs}\mkern-5mu≈\mkern-6mu\Varid{ys})\;{}\<[28]%
\>[28]{}\mathrel{=}\;\Varid{sound}_{\text{W}}\;\Varid{valid}\;\Varid{xs}\mkern-5mu≈\mkern-6mu\Varid{ys}{}\<[E]%
\\
\>[B]{}\Varid{whnf}\;(\constructor{trans}\;\Varid{xs}\mkern-5mu≈\mkern-6mu\Varid{ys}\;\Varid{ys}\mkern-5mu≈\mkern-6mu\Varid{zs})\;{}\<[28]%
\>[28]{}\mathrel{=}\;\Varid{trans}_{\text{W}}\;(\Varid{whnf}\;\Varid{xs}\mkern-5mu≈\mkern-6mu\Varid{ys})\;(\Varid{whnf}\;\Varid{ys}\mkern-5mu≈\mkern-6mu\Varid{zs}){}\<[E]%
\ColumnHook
\end{hscode}\resethooks
The remaining pieces of the soundness proof are omitted (see
\citet{accompanying-code-par-2010}).

\bibliographystyle{plainnat}
\bibliography{codata}

\end{document}